\documentclass[useAMS,usenatbib]{mn2e}
\usepackage{amsmath,natbib,lscape,graphicx,natbib,amssymb,times,multirow,epsfig}
\include{aas_macros}

\title[Companions to X-ray detected B6--A7 stars]{The VAST
  Survey -- I. Companions and the unexpected X-ray detection of B6--A7 stars}

\author[De Rosa et al.]
{R. J. De Rosa$^1$\thanks{E-mail: derosa@astro.ex.ac.uk}, J. Bulger$^1$, J. Patience$^{1,2}$, B. Leland$^1$,
  B. Macintosh$^2$, A. Schneider$^3$,\newauthor I. Song$^3$,
  C. Marois$^4$, J. R. Graham$^{5,6}$, M. Bessell$^7$ \& R. Doyon$^8$\\
$^1$ School of Physics, Exeter University, Stocker Road, Exeter, EX4 4QL, United
Kingdom\\
$^2$ Institute of Geophysics and Planetary Physics, Lawrence Livermore National Laboratory, 7000 East Ave, Livermore, CA 94550,
USA\\
$^3$ Physics and Astronomy, University of Georgia, 240 Physics, Athens, GA 30602,
USA\\
$^4$ NRC Herzberg Institute of Astrophysics, 5071 West Saanich Road, Victoria,
BC, V9E 2E7, Canada\\
$^5$ Astronomy Department, University of California, Berkeley, CA
94720, USA\\
$^6$ Dunlap Institute for Astronomy and Astrophysics, University of Toronto, 50 St. George Street, Toronto, ON, M55 3H8, Canada\\
$^7$ Mount Stromlo and Siding Spring Observatories, Institute of Advanced
Studies, The Australian National University, Weston Creek P.O., ACT 2611,
Australia\\
$^8$ D\'{e}pt de Physique, Universit\'{e} de Montr\'{e}al, C.P. 6128, Succ. Centre-Ville, Montr\'{e}al, QC,
H3C 3J7, Canada \\}
\pagerange{\pageref{firstpage}--\pageref{lastpage}} \pubyear{2010}
\begin{document}
\label{firstpage}

\maketitle

\begin{abstract}
With an adaptive optics imaging survey of 148 B6-A7 stars, we have
tested the hypothesis that unresolved lower-mass companions are the
source of the unexpected X-ray detections of stars in this spectral
type range. The sample is composed of 63 stars detected in X-rays within
the {\sl ROSAT} All-Sky Survey and 85 stars that form a control sample; both
subsets have the same restricted distribution of spectral type, age, X-ray
sensitivity and separation coverage. A total
of 68 companion candidates are resolved with separations
  ranging from 0$\farcs$3 to 26$\farcs$2, with 23 new
detections. The multiple star frequency of the X-ray sample based on
companions resolved within the {\sl ROSAT} error ellipse is found to be
$43^{+6}_{-6}\%$. The corresponding control sample multiple star
frequency is three times lower at $12^{+4}_{-3}\%$ -- a
difference of $31\pm7\%$. These results are presented in
the first of a series of papers based on our {\bf V}olume-limited {\bf
  A}-{\bf St}ar (VAST) survey -- a comprehensive study of the
multiplicity of A-type stars.

\end{abstract}

\section{Introduction}
The detection of X-ray emission from Main Sequence stars
is common \citep{Vaiana:1981p2184}, with the notable exception of late
B- and early A-type stars (e.g. \citealp{Stauffer:1994p2737}). Two
distinct generation mechanisms are responsible for the X-ray
emission, related to the different stellar structure of massive O- and
B-type stars and lower mass F- to M-type stars. For the massive stars, 
the hot stellar winds cause X-ray emission,
while the lower mass stars produce X-rays from the confinement of superheated 
plasma within their magnetic fields.

Radiative winds driven by line-absorption and re-emission within the
extended atmospheres of O- and early B-type stars form a key component
of the model for X-ray emission from these massive stars
(e.g. \citealp{Lucy:1980p1094}). Wind shocks caused either through
instability generated through  radiative driving
\citep{Owocki:1988p2431}, or due to collisions of magnetically driven
wind streams \citep{Feldmeier:1997p2436} are thought to be the primary
X-ray generation mechanisms. Interaction between stellar winds and
surrounding material is also thought to produce X-rays
(e.g. \citealp{Giampapa:1998p1715}). 

For lower mass stars,
stellar winds are too weak to generate X-rays, and the stellar corona is
responsible for the emission of X-rays and is intrinsically linked to
the magnetic field. For late A- to early M-type stars, magnetic fields arise
from the $\alpha\Omega$ dynamo caused by the differential
rotation at the interface between the convective envelope and the radiative core
\citep{Spiegel:1980p1537}. The magnetic
field generated by the dynamo process is essential for confining the
superheated plasma necessary for X-ray generation
\citep{Gudel:2004p1100}. The heating mechanism required to maintain
the corona at temperatures greater than $10^6$ K was originally thought
to be acoustic waves
(e.g. \citealp{Schwarzschild:1948p2101,Schatzman:1949p2105}), while
current models involve Alfv\'{e}n waves travelling 
perpendicular to the magnetic field
(e.g. \citealp{DePontieu:2007p2111,Jess:2009p2115}). Localised
magnetic reconnection events within the chromosphere are also
a potential source of coronal heating through Joule heating
(e.g. \citealp{Sturrock:1999p2661}). Beginning at a spectral type of
M3 ($\sim 0.35M_{\odot}$), the stars become fully convective
\citep{Chabrier:1997p2652} and the high level of
magnetic activity observed (e.g. \citealp{Randich:2000p2660}) may be
due to an $\alpha^2$-type dynamo generation mechanism
\citep{Chabrier:2006p1541}, in which turbulent motions are able to
generate large-scale magnetic fields.

In addition to the emission mechanisms intrinsic to the star, X-rays can be
generated by processes involving binary systems. Accretion of material
within cataclysmic variable systems
(e.g. \citealp{Patterson:1985p1492}) and compact object binaries
(e.g. \citealp{Shapiro:1976p1501}) can produce X-ray fluxes. For stars
between spectral types B6 to A7, which are expected to be X-ray quiet,
the presence of a low-mass companion can lead to the detection
of X-rays which are assigned to the primary if the companion is
unknown. This study is designed to explore the hypothesis that
unresolved lower mass companions are the true source of the unexpected X-ray
detections from B6-A7 stars.
\section{Previous observations}
\subsection{X-ray detection of B6-A7 stars}
Early studies of stellar X-ray emission conducted with the {\sl Einstein
  Observatory} measured a notable decrease in the fraction of X-ray
detected A-type stars ($0.00 \lesssim B-V \lesssim 0.25 $) compared to
bluer and redder stars \citep{Topka:1982p2188,Schmitt:1985p1699}. Out
of the 35 A-type stars observed by \citet{Schmitt:1985p1699}, only 7 were
detected and 4 were listed as having a secondary component which could
be the source of the X-ray emission. {\sl Einstein} observations of
coeval stellar groups also showed a similar decrease in the fraction
of X-ray detections of A-type stars between $0.00 \lesssim B-V \lesssim 0.3
$ (e.g. \citealp{Micela:1985p1704,Schmitt:1990p1707}).

The increased sensitivity provided by the {\sl ROSAT} mission and
all-sky coverage led to the detection and characterisation of a
significant number stellar X-ray sources
\citep{Voges:1999p197}. A search by \citet{Huensch:1998p709} of the
Bright Star Catalogue \citep{Hoffleit:1964p2777} and the {\sl ROSAT} Bright
Source Catalogue for objects within 90$\arcsec$ of the same position
defined a population of 232 B6-A7 X-ray detected stars. To investigate
possible sources of the X-ray emission for this sample, the X-ray
luminosity was compared with spectral type, spectral peculiarities and
rotational velocities (e.g. \citealp{Simon:1995p917,Schroder:2007p864}). The lack of a
dependence on any of these factors was taken as evidence of unresolved
companions. Without a comprehensive binary survey of A-type stars, it was
not possible to test the companion hypothesis directly. Similarly,
X-ray data from {\sl Chandra} which could resolve the emission
source in tight ($\rho\sim0\farcs5$) binary systems does not exist for a
significant sample of X-ray B6-A7 stars.
\subsection{High-resolution imaging companion searches}
High resolution AO imaging studies of X-ray detected B- and A-type stars
have been employed to search for lower mass stars capable of producing
X-rays. Pointed observations of late B-type stars with known lower
mass companions (e.g. \citealp{Schmitt:1993p745,Berghofer:1994p1709})
wide enough to be resolved with the {\sl ROSAT} High Resolution Imager were
obtained to determine the source of the X-ray emission. These
observations typically identified the B-type star as the source of X-ray
emission, although subsequent high-resolution AO imaging has revealed
additional components to several of these systems
(e.g. \citealp{Shatsky:2002p1710}). Sub-arcsecond binary companions
have also been resolved with high-resolution AO imaging
of pre-Main Sequence companions to late B-type stars
(e.g. \citealp{Hubrig:2001p1711}).

Recent discoveries of low-mass
companions to Alcor \citep{Mamajek:2010p722,Zimmerman:2010p727} and
$\zeta$ Virginis \citep{Hinkley:2010p1214} have both noted that the
unexplained X-ray emission from the primary can be explained by the
lower-mass companion, and demonstrate how  X-ray emission from A-type stars
could be a useful tool in searching for low-mass companions. The
current study expands upon the existing imaging results of  X-ray
detected B6-A7 stars by observing a large sample of both X-ray stars
and a control sample.
\section{Sample}
\begin{table*}
\caption{X-Ray Detected Sample}
{\tiny \begin{tabular}{ccccccccccccc}
HIP&Name&{\sl
  Hipparcos}&Distance&\multicolumn{3}{c}{{\sl ROSAT} Source}&\multicolumn{2}{c}{Observations}&Magnitude&Band&Integration&2MASS\\
&&Spectral&&Designation&Offset&Error Radius&Tel.&Date&&&&Sources\\
&&Type&\textit{pc}&B - BSC, F - FSC&\textit{arcsec}&\textit{arcsec}&&&&&\textit{sec}&\textit{arcmin$^{-2}$}\\
\hline
5310&$\psi^2$ Psc&A3V&49.4$\pm$2.0&B - J010757.4+204424&4.9&14&Gemini&16/10/2008&5.22$\pm$0.02&K&200&0.156\\
9480&48 Cas&A3IV&35.8$\pm$0.7&B - J020156.9+705432&7.6&8&CFHT&01/09/2009&4.25$\pm$0.27&K&480&0.868\\
11569&$\iota$ Cas&A5p&43.4$\pm$1.5&B - J022902.9+672407&6.4&8&CFHT&05/02/2010&4.25$\pm$0.03&K&330&1.088\\
13133&RZ Cas&A3Vv+&62.5$\pm$2.4&B - J024854.7+693804&4.3&7&Gemini&14/11/2008&5.47$\pm$0.02&K&400&0.767\\
17608&Merope&B6IVe&110.1$\pm$12.6&F - J034620.7+235713&24.5&13&AEOS&04/02/2002&4.14$\pm$0.03&I&300&0.207\\
17664&&B9.5V&150.1$\pm$22.3&F - J034659.4+243049&23.5&23&AEOS&02/03/2003&6.79$\pm$0.01&I&599&0.216\\
17923&&A0V&158.7$\pm$36.8&B - J034958.2+235109&13.8&14&AEOS&03/02/2002&6.74$\pm$0.01&I&289&0.212\\
19949&&A2Vn&108.2$\pm$8.8&F - J041642.9+533649&6.7&16&AEOS&05/02/2002&5.15$\pm$0.00&I&300&0.749\\
20070&b Per&A2V&97.6$\pm$8.3&B - J041814.8+501747&3.7&7&AEOS&05/02/2002&4.44$\pm$0.03&I&300&0.800\\
20156&&A7V&79.4$\pm$5.3&B - J041913.6+500254&3.6&8&AEOS&05/02/2002&5.23$\pm$0.01&I&300&0.781\\
20380&&A3V&95.0$\pm$6.1&F - J042149.8+563020&16.8&27&AEOS&05/02/2002&5.79$\pm$0.00&I&300&0.677\\
20400&60 Tau&A3m&45.7$\pm$2.0&F - J042204.5+140440&14.7&27&AEOS&04/02/2002&5.38$\pm$0.03&I&300&0.150\\
20484&63 Tau&A1m&47.2$\pm$1.8&F - J042325.5+164633&8.1&28&AEOS&04/02/2002&5.33$\pm$0.03&I&300&0.180\\
20648&$\delta^3$ Tau&A2IV&45.3$\pm$1.6&F - J042528.4+175512&31.8&14&AEOS&04/02/2002&4.24$\pm$0.03&I&300&0.189\\
&&&&&&&CFHT&04/02/2010&4.10$\pm$0.03&K&352&0.317\\
21402&88 Tau&A5m&46.1$\pm$1.7&B - J043538.5+100941&11.6&8&AEOS&04/02/2002&4.06$\pm$0.02&I&300&0.171\\
22287&4 Cam&A3m&49.7$\pm$2.0&F - J044758.6+564531&14.6&18&AEOS&05/02/2002&5.08$\pm$0.03&I&300&0.506\\
23040&7 Cam&A1V&115.2$\pm$10.8&B - J045714.3+534442&34.8&10&AEOS&05/02/2002&4.39$\pm$0.03&I&300&0.545\\
23179&$\omega$ Aur&A1V&48.8$\pm$2.2&B - J045915.4+375330&5.2&7&Gemini&15/11/2008&4.92$\pm$0.03&K&200&1.125\\
23875&$\beta$ Eri&A4III&27.2$\pm$0.6&B - J050750.5-050455&17.3&10&Gemini&19/12/2009&2.40$\pm$0.22&K&200&0.274\\
24019&&A5m&54.7$\pm$3.9&F - J050945.2+280209&20&12&AEOS&04/02/2002&5.69$\pm$0.01&I&300&0.466\\
26126&38 Ori&A2V&105.8$\pm$8.3&F - J053416.4+034623&23.4&17&AEOS&05/02/2002&5.25$\pm$0.03&I&300&0.322\\
28614&$\mu$ Ori&A1Vm&46.5$\pm$1.8&B - J060222.9+093854&4.6&9&Gemini&19/12/2009&3.64$\pm$0.26&K&200&0.995\\
29997&&A0Vn&53.9$\pm$1.9&F - J061849.6+691929&18.7&14&CFHT&01/09/2009&4.67$\pm$0.02&K&480&0.310\\
30060&2 Lyn&A2Vs&45.7$\pm$2.0&F - J061938.6+590019&22.4&20&AEOS&05/02/2002&4.40$\pm$0.00&I&300&0.243\\
&&&&&&&CFHT&01/09/2009&4.35$\pm$0.02&K&480&0.364\\
30419&$\epsilon$ Mon&A5IV&39.4$\pm$1.6&B - J062346.2+043544&9.9&10&CFHT&01/09/2009&3.92$\pm$0.04&K&480&1.133\\
35643&&A7s&34.5$\pm$0.9&F - J072119.0+451327&20.9&22&AEOS&02/02/2002&5.38$\pm$0.01&I&300&0.183\\
39095&&A1V&73.1$\pm$4.3&F - J075951.6-182353&7.4&17&AEOS&02/02/2002&4.51$\pm$0.02&I&300&0.756\\
39847&27 Lyn&A2V&66.8$\pm$3.1&F - J080828.5+513040&19.3&16&AEOS&02/02/2002&4.71$\pm$0.04&I&300&0.126\\
41564&&A5m&85.2$\pm$6.8&B - J082828.5-023051&13.9&9&AEOS&03/02/2002&6.06$\pm$0.01&I&300&0.235\\
42313&$\delta$ Hya&A1Vnn&54.9$\pm$2.7&B - J083740.1+054217&11.2&11&AEOS&04/02/2002&4.12$\pm$0.02&I&300&0.166\\
&&&&&&&AEOS&01/03/2003&4.12$\pm$0.02&I&300&0.166\\
44127&$\iota$ Uma&A7V&14.6$\pm$0.2&B - J085913.0+480227&5.9&11&Palomar&12/04/2008&2.66$\pm$0.24&K&71&0.145\\
&&&&&&&CFHT&05/02/2010&2.66$\pm$0.24&K&352&0.145\\
45688&38 Lyn&A3V&37.4$\pm$1.1&B - J091850.2+364814&7.5&8&AEOS&06/02/2002&3.81$\pm$0.03&I&300&0.081\\
&&&&&&&Palomar&12/04/2008&3.42$\pm$0.35&K&71&0.124\\
51200&&A2V&66.3$\pm$3.1&F - J102728.3+413613&9.7&13&CFHT&04/02/2010&5.53$\pm$0.02&K&440&0.104\\
52913&40 Sex&A2IV&95.9$\pm$10.8&F - J104917.1-040123&4.7&14&AEOS&02/03/2003&6.38$\pm$0.01&I&300&0.083\\
57646&&A3m&62.7$\pm$3.2&F - J114915.1+161430&6.4&12&AEOS&02/03/2003&5.75$\pm$0.01&I&300&0.058\\
&&&&&&&CFHT&05/02/2010&5.35$\pm$0.02&K&352&0.091\\
58001&Phecda&A0Ve&25.6$\pm$0.4&B - J115352.3+534153&25&16&Palomar&11/04/2008&2.49$\pm$0.17&H&71&0.090\\
&&&&&&&CFHT&05/02/2010&2.43$\pm$0.29&K&352&0.102\\
59504&&A5m&33.7$\pm$0.6&B - J121210.1+773702&7.3&20&AEOS&29/05/2002&4.77$\pm$0.00&I&300&0.103\\
62394&34 Vir&A3V&74.6$\pm$5.2&F - J124714.4+115723&12.5&19&AEOS&29/05/2002&5.98$\pm$0.00&I&300&0.059\\
62572&&A1IIIshe&93.0$\pm$15.1&F - J124909.8+832448&6.7&13&AEOS&29/05/2002&5.26$\pm$0.00&I&300&0.126\\
65198&&A2V&65.3$\pm$3.4&B - J132141.7+020521&7.2&8&AEOS&02/03/2003&5.64$\pm$0.00&I&300&0.075\\
&&&&&&&CFHT&13/06/2008&5.60$\pm$0.05&H&450&0.103\\
&&&&&&&CFHT&04/02/2010&5.51$\pm$0.02&K&440&0.115\\
65241&64 Vir&A2m&63.7$\pm$2.9&B - J132209.8+050918&2.6&10&AEOS&03/03/2003&5.82$\pm$0.03&I&319&0.066\\
&&&&&&&CFHT&14/06/2008&5.67$\pm$0.03&H&450&0.094\\
65477&Alcor&A5V&24.9$\pm$0.3&F - J132513.8+545920&3.9&13&Palomar&11/04/2008&3.30$\pm$0.23&H&71&0.100\\
66249&$\zeta$ Vir&A3V&22.5 $\pm$ 0.4&F - J133442.6-003530&19.6&14&AEOS&03/03/2003&3.27$\pm$0.02&I&300&0.079\\
&&&&&&&CFHT&05/02/2010&3.22$\pm$0.27&K&352&0.122\\
66727&1 Boo&A1V&92.8$\pm$7.8&B - J134040.2+195708&12.8&11&AEOS&03/03/2003&5.72$\pm$0.01&I&300&0.064\\
71618&33 Boo&A1V&60.4$\pm$2.0&B - J143850.0+442418&3.7&7&CFHT&14/06/2008&5.28$\pm$0.04&H&450&0.098\\
76376&&A2V&75.5$\pm$3.0&B - J153556.8+54375&3.3&9&AEOS&29/05/2002&5.72$\pm$0.00&I&300&0.089\\
76878&$\tau^7$ Ser&A2m&53.2$\pm$2.3&B - J154154.9+182744&6.9&15&AEOS&29/05/2002&5.61$\pm$0.03&I&300&0.096\\
&&&&&6.9&15&Palomar&13/07/2008&5.30$\pm$0.02&K&62&0.149\\
77336&$\upsilon$ Ser&A3V&77.2$\pm$6.0&F - J154717.7+140652&7.3&16&AEOS&29/05/2002&5.61$\pm$0.00&I&300&0.108\\
80628&$\upsilon$ Oph&A3m&37.5$\pm$1.2&B - J162748.2-082213&4.8&9&Palomar&12/04/2008&4.17$\pm$0.04&K&47&0.401\\
82321&52 Her&A2Vspe&53.7$\pm$1.5&F - J164914.1+455848&12&12&Palomar&12/07/2008&4.58$\pm$0.04&H&71&0.174\\
83223&&A7V&73.1$\pm$4.6&F - J170028.6+063456&13.4&29&AEOS&29/05/2002&6.33$\pm$0.01&I&300&0.232\\
85829&$\nu^2$ Dra&A4m&30.6$\pm$0.5&F - J173216.1+551023&1&15&Palomar&12/04/2008&4.16$\pm$0.02&K&71&0.203\\
&&&&&&&CFHT&05/02/2010&4.16$\pm$0.02&K&264&0.223\\
87045&&A2Vs&131.6$\pm$14.2&B - J174707.6+473648&5.2&8&AEOS&29/05/2002&6.37$\pm$0.01&I&300&0.183\\
87212&30 Dra&A2V&66.5$\pm$2.1&F - J174904.5+504651&2.6&34&Gemini&24/06/2008&4.88$\pm$0.02&K&200&0.259\\
&&&&&&&Palomar&12/07/2008&4.88$\pm$0.02&K&71&0.259\\
88771&72 Oph&A4IVs&25.4$\pm$0.5&F - J180719.8+093411&26.4&16&Palomar&12/04/2008&3.41$\pm$0.19&K&71&0.990\\
&&&&&&&CFHT&05/02/2010&3.41$\pm$0.19&K&264&0.990\\
89925&108 Her&A5m&57.6 $\pm$ 2.0&F - J182057.4+295146&14.8&12&AEOS&31/05/2002&5.37$\pm$0.03&I&300&0.348\\
&&&&&&&Gemini&24/06/2008&4.99$\pm$0.02&K&200&0.540\\
&&&&&&&CFHT&01/09/2009&4.99$\pm$0.02&K&720&0.540\\
91971&$\zeta^1$ Lyr&Am&47.1$\pm$1.2&B - J184446.1+373620&3.7&9&CFHT&13/06/2008&3.97$\pm$0.23&K&180&0.566\\
93747&$\zeta$ Aql&A0Vn&25.5$\pm$0.5&B - J190526.0+135136&22.8&12&CFHT&13/06/2008&3.05$\pm$0.28&H&180&5.932\\
98103&$\phi$ Aql&A1IV&63.1$\pm$3.0&F - J195613.8+112526&8.5&13&Gemini&18/06/2008&5.26$\pm$0.02&K&200&1.778\\
102033&&A2V&82.0$\pm$4.4&F - J204036.8+294822&6.1&15&AEOS&31/05/2002&5.87$\pm$0.00&I&300&0.993\\
106711&74 Cyg&A5V&63.3$\pm$2.5&F - J213656.7+402440&8.9&14&Gemini&08/09/2008&4.51$\pm$0.02&K&200&1.218\\
109521&&A5V&54.8$\pm$1.7&B - J221109.0+504929&8.7&14&Gemini&08/09/2008&4.96$\pm$0.02&K&200&2.059\\
110787&$\rho^1$ Cep&A2m&62.6$\pm$2.0&F - J22264.9+784709&4.5&12&Gemini&17/08/2008&5.54$\pm$0.03&K&200&0.435\\
117452&$\delta$ Scl&A0V&44.0$\pm$2.2&B - J234854.7-280751&11.2&13&CFHT&30/08/2009&4.53$\pm$0.02&K&480&0.112\\
\label{tab:sample}
\end{tabular}}
\end{table*}
\begin{table*}
\caption{Control Sample}
{\tiny \begin{tabular}{cccccccccc}
HIP&Name&{\sl Hipparcos}&Distance&\multicolumn{2}{c}{Observations}&Magnitude&Band&Integration&2MASS
Sources\\
&&Spectral Type&\textit{pc}&Tel.&Date&&&\textit{sec}&\textit{arcmin$^{-2}$}\\
\hline
159&&A3&59.1$\pm$2.8&Gemini&17/10/2008&6.21$\pm$0.02&K&200&0.115\\
2852&&A5m&49.7$\pm$2.2&Gemini&17/10/2008&5.42$\pm$0.02&K&200&0.100\\
3414&$\pi$ Cas&A5V&53.5$\pm$2.1&AEOS&02/03/2003&4.79$\pm$0.01&I&300&0.334\\
&&&&Gemini&16/10/2008&4.58$\pm$0.02&K&200&0.505\\
5317&41 And&A3m&60.2$\pm$2.7&Gemini&16/10/2008&4.77$\pm$0.02&K&200&0.404\\
8122&&A3&71.7$\pm$3.8&CFHT&30/08/2009&6.17$\pm$0.02&K&640&0.229\\
9487&$\alpha$ Psc&A2&42.6$\pm$1.9&CFHT&01/09/2009&3.62$\pm$0.33&K&480&0.115\\
13717&&A1Vn&57.9$\pm$3.1&Gemini&18/10/2008&4.86$\pm$0.02&K&200&0.115\\
17489&Celeno&B7IV&102.6$\pm$11.1&AEOS&03/02/2002&5.43$\pm$0.03&I&300&0.207\\
17572&&A0V&103.3$\pm$11.0&AEOS&04/02/2002&6.77$\pm$0.01&I&300&0.192\\
17588&22 Tau&A0Vn&108.6$\pm$10.9&AEOS&02/02/2002&6.41$\pm$0.03&I&300&0.213\\
17791&&A1V&144.9$\pm$20.8&AEOS&04/02/2002&6.83$\pm$0.02&I&300&0.217\\
17847&Atlas&B8III&116.7$\pm$14.0&AEOS&04/02/2002&3.64$\pm$0.03&I&300&0.214\\
20507&$\xi$ Eri&A2V&63.9$\pm$3.3&Gemini&17/10/2008&4.93$\pm$0.02&K&200&0.187\\
20641&$\kappa^2$ Tau&A7&44.2$\pm$1.6&CFHT&05/02/2010&4.61$\pm$0.02&K&352&0.362\\
20894&$\theta^2$ Tau&A7III&45.7$\pm$1.7&CFHT&04/02/2010&2.88$\pm$0.26&K&440&0.307\\
21039&81 Tau&Am&44.3$\pm$2.1&CFHT&05/02/2010&4.90$\pm$0.02&K&352&0.311\\
22192&EX Eri&A3IV&57.5$\pm$2.2&Gemini&14/11/2008&5.72$\pm$0.02&K&200&0.172\\
23554&&A2IV&60.1$\pm$2.3&Gemini&19/12/2009&5.34$\pm$0.02&K&200&0.209\\
23983&16 Ori&A2m&53.9$\pm$2.4&Gemini&05/11/2008&4.86$\pm$0.02&K&200&0.403\\
25197&16 Cam&A0Vne&104.3$\pm$8.4&AEOS&05/02/2002&5.23$\pm$0.00&I&300&0.337\\
26309&&A2III-IV&56.6$\pm$2.3&Gemini&14/11/2008&5.86$\pm$0.02&K&200&0.255\\
28360&Menkalinan&A2IV+&25.2$\pm$0.5&CFHT&05/02/2010&1.78$\pm$0.19&K&352&0.651\\
28910&$\theta$ Lep&A0V&52.2$\pm$1.9&Gemini&25/11/2008&4.52$\pm$0.02&K&200&0.472\\
29711&&A5IVs&66.5$\pm$3.4&Gemini&25/11/2008&5.86$\pm$0.02&K&200&0.729\\
31119&&A3V&64.8$\pm$3.6&AEOS&04/02/2002&5.04$\pm$0.01&I&300&0.706\\
&&&&Gemini&11/11/2008&4.77$\pm$0.04&K&200&1.198\\
31290&&A3V&136.1$\pm$18.0&AEOS&05/02/2002&6.46$\pm$0.01&I&300&0.402\\
34897&&A5&66.4$\pm$3.4&Gemini&10/05/2010&5.99$\pm$0.02&K&200&0.293\\
35341&65 Aur&A5Vn&82.1$\pm$6.2&AEOS&02/02/2002&5.69$\pm$0.01&I&300&0.207\\
35350&$\lambda$ Gem&A3V&28.9$\pm$0.8&Palomar&12/04/2008&3.54$\pm$0.26&K&68&0.489\\
&&&&CFHT&04/02/2010&3.54$\pm$0.26&K&440&0.489\\
38723&&A3p&60.4$\pm$3.4&Gemini&11/11/2008&5.40$\pm$0.02&K&200&0.241\\
40646&29 Lyn&A7IV&93.2$\pm$5.9&AEOS&03/02/2002&5.46$\pm$0.00&I&300&0.116\\
41152&&A3V&51.4$\pm$1.9&AEOS&06/02/2002&5.39$\pm$0.01&I&300&0.113\\
42806&Asellus Borealis&A1IV&48.6$\pm$2.0&AEOS&02/03/2002&4.64$\pm$0.00&I&300&0.123\\
43570&&A5V&167.8$\pm$38.8&AEOS&04/02/2002&6.21$\pm$0.01&I&300&0.138\\
43932&$\sigma^2$ Cnc&A7IV&59.8$\pm$3.3&AEOS&06/02/2002&5.26$\pm$0.01&I&300&0.099\\
44066&$\alpha$ Cnc&A5m&53.2$\pm$2.8&AEOS&01/03/2003&4.13$\pm$0.03&I&300&0.119\\
44901&15 UMa&A1m&29.3$\pm$0.7&AEOS&01/03/2003&4.19$\pm$0.03&I&300&0.087\\
&&&&Palomar&12/04/2008&4.04$\pm$0.28&K&69&0.138\\
&&&&CFHT&05/02/2010&4.04$\pm$0.28&K&352&0.138\\
45493&18 UMa&A5V&36.3$\pm$1.0&AEOS&03/03/2003&4.58$\pm$0.03&I&300&0.094\\
&&&&Palomar&12/04/2008&4.29$\pm$0.02&K&85&0.138\\
49593&21 LMi&A7V&28.0$\pm$0.7&Palomar&12/04/2008&4.00$\pm$0.04&K&44&0.108\\
51658&&A7IV&34.3$\pm$0.9&Palomar&12/04/2008&4.20$\pm$0.02&K&69&0.106\\
&&&&CFHT&04/02/2010&4.20$\pm$0.02&K&440&0.106\\
53910&Merak&A1V&24.4$\pm$0.4&CFHT&04/02/2010&2.29$\pm$0.24&K&440&0.106\\
53954&60 Leo&A1m&37.9$\pm$1.2&Palomar&12/04/2008&4.32$\pm$0.04&K&71&0.103\\
54063&&A5&61.8$\pm$3.3&CFHT&14/06/2008&6.29$\pm$0.02&H&450&0.079\\
54136&51 UMa&A3III-IV&80.6$\pm$4.6&AEOS&03/03/2003&5.86$\pm$0.01&I&300&0.065\\
57328&$\xi$ Vir&A4V&36.6$\pm$1.1&AEOS&03/03/2003&4.67$\pm$0.00&I&300&0.067\\
&&&&Palomar&11/04/2008&4.41$\pm$0.05&K&71&0.104\\
&&&&CFHT&13/06/2008&4.54$\pm$0.08&H&180&0.091\\ 
57632&$\beta$ Leo&A3V&11.1 $\pm$ 0.1&CFHT&05/02/2010&1.88$\pm$0.19&K&352&0.099\\
58510&7 Vir&A1V&84.8$\pm$5.4&AEOS&02/03/2003&5.34$\pm$0.00&I&290&0.068\\
58590&$\pi$ Vir&A5V&109.2$\pm$10.1&AEOS&02/03/2003&4.53$\pm$0.02&I&300&0.067\\
59394&3 Crv&A1V&56.1$\pm$2.1&CFHT&14/06/2008&5.36$\pm$0.04&H&450&0.165\\
59608&12 Vir&A2m&49.5$\pm$1.8&AEOS&03/03/2003&5.61$\pm$0.03&I&200&0.060\\
&&&&CFHT&04/02/2010&5.24$\pm$0.02&K&440&0.097\\
59774&Megrez&A3V&25.0$\pm$0.4&Palomar&11/04/2008&3.31$\pm$0.25&H&71&0.094\\
&&&&CFHT&13/06/2008&3.31$\pm$0.25&H&180&0.094\\
60746&16 Com&A4V&86.5$\pm$5.6&AEOS&03/02/2003&4.93$\pm$0.03&I&300&0.058\\
61960&$\rho$ Vir&A0V&36.9$\pm$1.1&AEOS&02/03/2003&4.80$\pm$0.03&I&300&0.064\\
&&&&Palomar&11/04/2008&4.76$\pm$0.02&H&57&0.092\\
&&&&CFHT&04/02/2010&4.68$\pm$0.02&K&440&1.040\\
62933&41 Vir&A7III&61.0$\pm$2.9&CFHT&05/02/2010&5.47$\pm$0.02&K&352&0.097\\
68520&$\tau$ Vir&A3V&66.9$\pm$3.9&AEOS&03/03/2003&4.11$\pm$0.02&I&300&0.075\\
69592&&A7V&59.0$\pm$2.6&Palomar&12/07/2008&5.90$\pm$0.02&H&71&0.093\\
69732&$\lambda$ Boo&A0p&29.8$\pm$0.5&AEOS&03/03/2003&4.15$\pm$0.03&I&300&0.069\\
&&&&Palomar&11/04/2008&4.03$\pm$0.25&H&71&0.095\\
69951&&A5&73.5$\pm$3.4&Palomar&12/07/2008&6.40$\pm$0.04&H&71&0.097\\
69974&$\lambda$ Vir&A1V&57.2$\pm$3.1&AEOS&03/03/2003&4.43$\pm$0.02&I&669&0.114\\
&&&&Gemini&24/06/2008&4.24$\pm$0.02&K&200&0.177\\
71075&$\gamma$ Boo&A7III&26.1$\pm$0.5&Palomar&11/04/2008&2.57$\pm$0.25&H&71&0.089\\
75043&&A4V&65.3$\pm$2.2&AEOS&29/05/2002&5.52$\pm$0.00&I&300&0.082\\
76852&$\iota$ Ser&A1V&58.9$\pm$2.7&Palomar&13/07/2008&4.31$\pm$0.02&K&71&0.142\\
77233&$\beta$ Ser&A3V&46.9$\pm$1.9&Palomar&12/04/2008&3.55$\pm$0.32&K&67&0.161\\
77464&&A5IV&49.2$\pm$1.7&Gemini&27/06/2008&5.26$\pm$0.02&K&200&0.238\\
77622&$\epsilon$ Ser&A2m&21.6$\pm$0.3&Palomar&12/04/2007&3.43$\pm$0.27&K&71&0.203\\
83613&60 Her&A4IV&44.1$\pm$1.4&Palomar&12/04/2008&4.61$\pm$0.02&K&71&0.347\\
84012&$\eta$ Oph&A2IV-V&25.8$\pm$0.6&Palomar&12/04/2008&2.34$\pm$0.24&K&71&1.436\\
86565&o Ser&A2Va&51.5$\pm$2.8&Gemini&28/06/2008&4.11$\pm$0.25&K&200&3.249\\
87108&$\gamma$ Oph&A0V&29.0$\pm$0.8&Palomar&12/04/2008&3.62$\pm$0.23&K&71&0.919\\
&&&&CFHT&31/08/2009&3.62$\pm$0.23&K&480&0.919\\
92161&111 Her&A5III&28.4$\pm$0.6&CFHT&13/06/2008&4.08$\pm$0.03&H&180&1.472\\
95081&$\pi$ Dra&A2IIIs&68.9$\pm$2.2&Gemini&24/06/2008&4.45$\pm$0.02&K&200&0.380\\
&&&&Palomar&12/07/2008&4.58$\pm$0.17&H&71&0.348\\
95853&$\iota^2$ Cyg&A5Vn&37.5$\pm$0.6&CFHT&12/06/2008&3.69$\pm$0.23&H&280&0.583\\
99655&33 Cyg&A3IV-Vn&46.7$\pm$1.0&CFHT&14/06/2008&4.17$\pm$0.27&H&450&0.767\\
99742&$\rho$ Aql&A2V&47.1$\pm$1.7&CFHT&30/08/2009&4.77$\pm$0.02&K&480&1.353\\
99770&29 Cyg&A2V&41.0$\pm$0.9&CFHT&31/08/2009&4.42$\pm$0.02&K&480&7.030\\
100108&36 Cyg&A2V&59.7$\pm$1.9&AEOS&31/05/2002&5.51$\pm$0.00&I&300&2.521\\
&&&&Gemini&08/09/2009&5.49$\pm$0.02&K&200&6.666\\
100526&&A2&69.2$\pm$2.2&Gemini&08/09/2008&6.20$\pm$0.02&K&200&1.137\\
101093&$\theta$ Cep&A7III&41.6$\pm$0.9&CFHT&30/08/2009&3.72$\pm$0.32&K&200&0.662\\
101300&&Am&81.1$\pm$4.6&AEOS&31/05/2002&6.08$\pm$0.01&I&300&0.980\\
101483&$\eta$ Del&A3IVs&53.0$\pm$2.3&Gemini&08/09/2008&5.24$\pm$10.00&K&200&0.796\\
105966&35 Vul&A1V&55.7$\pm$2.2&Gemini&08/09/2008&5.29$\pm$0.02&K&200&0.630\\
109427&$\theta$ Peg&A1Va&29.6$\pm$0.8&CFHT&14/06/2008&3.39$\pm$0.21&H&450&0.193\\
109667&&A3V&58.1$\pm$2.7&Gemini&10/09/2008&5.74$\pm$0.02&K&200&0.157\\
&&&&CFHT&31/08/2009&5.74$\pm$0.02&K&720&0.157\\
111169&$\alpha$ Lac&A1V&31.4$\pm$0.5&Palomar&12/07/2008&3.87$\pm$0.21&H&71&1.384\\
&&&&CFHT&30/08/2009&3.85$\pm$0.27&K&480&1.516\\
116354&15 And&A1III&71.6$\pm$3.2&Gemini&08/09/2008&5.28$\pm$0.02&K&200&0.405\\
\label{tab:control}
\end{tabular}}
\end{table*}
\label{sec:sample}
\begin{figure}
\resizebox{\hsize}{!}{{\includegraphics{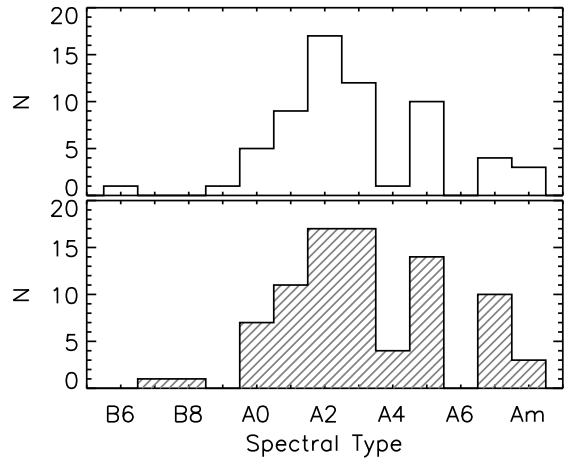}}} 
\caption{Distribution of the spectral type for each target reported
  wtithin the {\sl Hipparcos} catalogue for the X-ray (white
  histogram) and control (grey histogram) samples.}
\label{fig1}
\end{figure}
\begin{figure}
\resizebox{\hsize}{!}{{\includegraphics{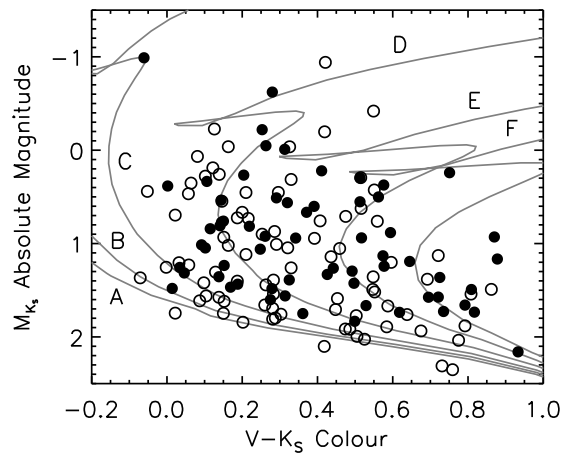}}} 
\caption{A colour-magnitude diagram of the X-ray (filled circles) and
  control (open circles) samples. Theoretical isochrones are plotted
  for A - 10Myrs, B - 100 Myrs, C - 250 Myrs, D - 500 Myrs, E - 800
  Myrs, F - 1 Gyrs \citep{Marigo:2008p2100}.}
\label{fig2}
\end{figure}
\begin{figure}
\resizebox{\hsize}{!}{{\includegraphics{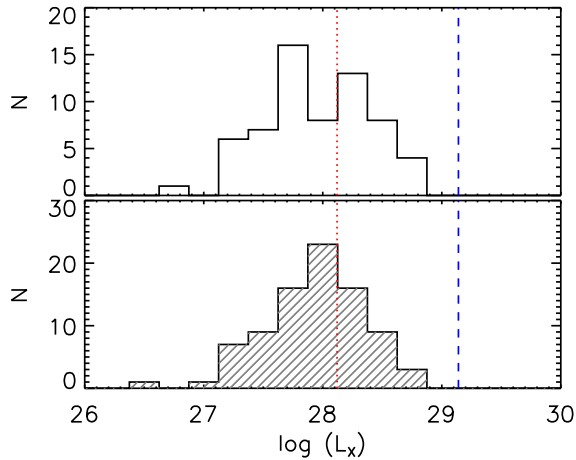}}}
\caption{Histogram of the RASS detection limits for the targets
  within the X-ray (white histogram) and control (grey histogram)
  samples. Mean $L_{\rm x}$ values for Pleiades (100 Myr) and
  Hyades (650 Myr) M-dwarfs as blue dashed and red dotted lines respectively
  \citep{Micela:1996p1713,Stern:1995p2131}.}
\label{fig3}
\end{figure}
\begin{figure}
\resizebox{\hsize}{!}{{\includegraphics{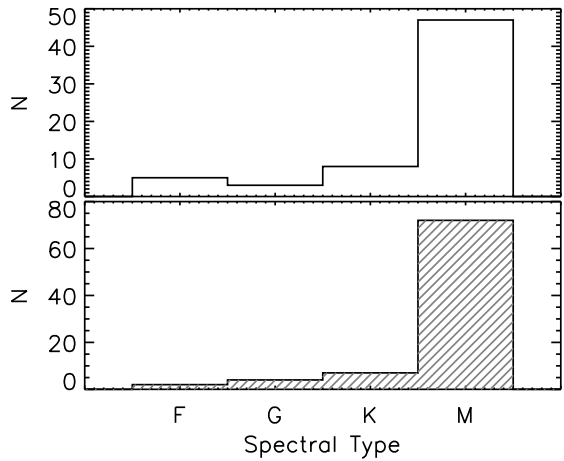}}}
\caption{Histogram showing the distribution of the sensitivity of RASS
  obserations to lower-mass companions for targets in the X-ray (white histogram) and control
  (grey histogram) samples. The evolution of $L_{\rm X}$ as a function
of stellar age was derived from observations of open clusters \citep[and references therein]{Gudel:2004p1100}.}
\label{fig4}
\end{figure}
\begin{figure}
\resizebox{\hsize}{!}{{\includegraphics{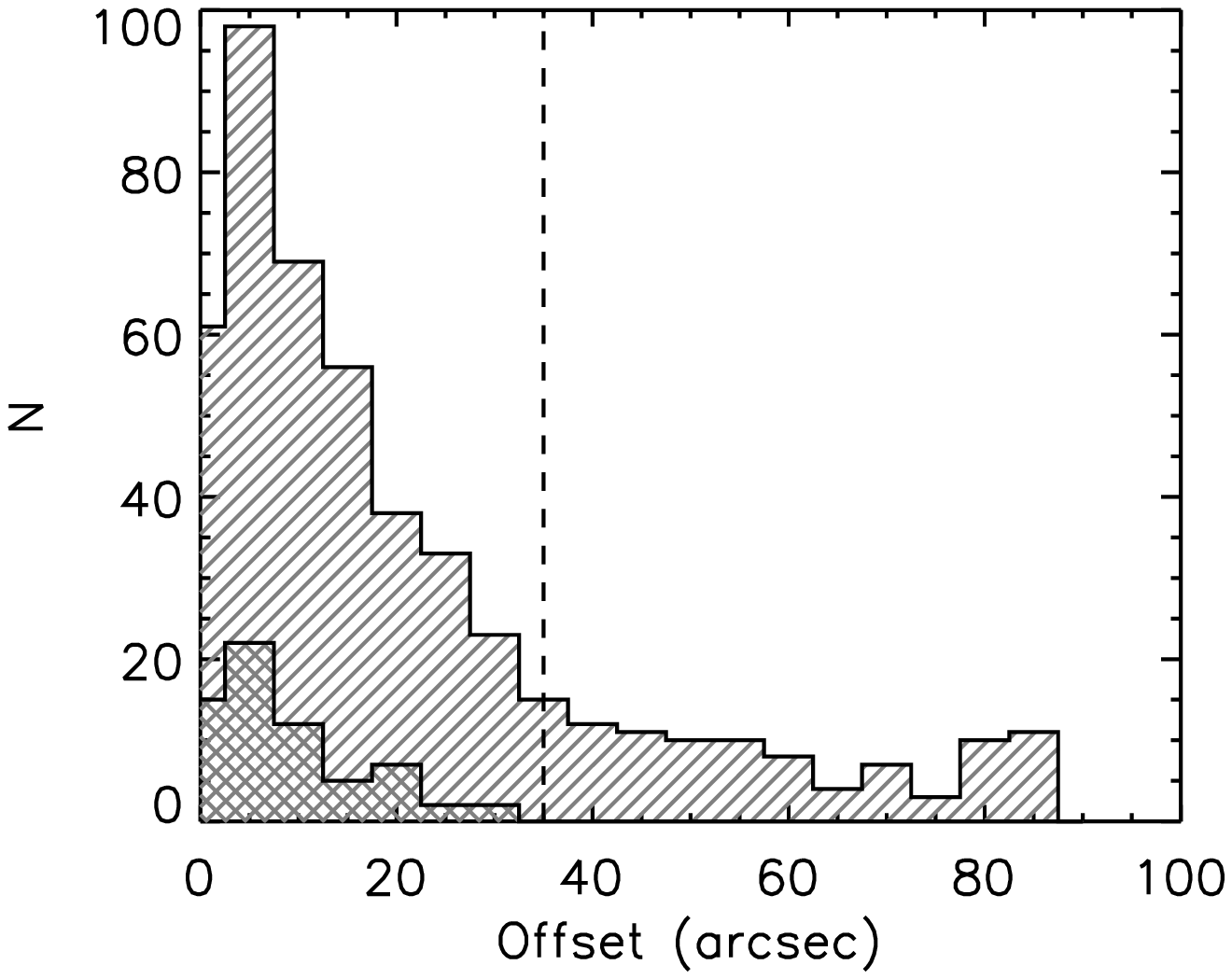}}}
\caption{Distribution of {\sl Hipparcos} -- {\sl ROSAT} offsets for
  479 X-ray detected early-type stars (hatched histogram). These stars have a {\sl
    ROSAT} source within $90\arcsec$ using the same criteria 
  as \citet{Huensch:1998p2050}. The tail of the distribution was
  removed by selecting a more stringent maximum offset of $35\arcsec$
  (dashed line). The offset distribution for the 63 early-type stars
  within the X-ray detected sample is overplotted (cross-hatched histogram).}
\label{fig5}
\end{figure}
Two samples were constructed in order to test the companion hypothesis:
a 63-star X-ray detected sample, and an 85-star control
sample. The distributions of spectral types reported in the {\sl
  Hipparcos} catalogue for both samples are shown in
Figure \ref{fig1}, and a K-S test confirms that both are drawn
from the same distribution. The majority of the total sample, 108
targets, form a part of our ongoing {\bf V}olume-limited {\bf A}-{\bf
  St}ar (VAST) survey which will include all A-type stars within 75
parsecs. Both the X-ray and control samples includes targets spanning a 
similar range of ages, as shown in the colour-magnitude diagram in
Figure \ref{fig2}. To perform a robust test of the companion
hypothesis, we ensured that each sample had a similar distribution of
sensitivity to X-ray sources. Background X-ray counts were extracted
from the {\sl ROSAT} All Sky Survey (RASS) observations at the
coordinates of each target within both samples. A minimum detectable
X-ray flux at each coordinate was estimated as five times the
background level. These minimum fluxes were calculated
assuming a hardness ratio of 0.5, typical of low-mass stellar sources
(e.g. \citealp{Huelamo:2000p2209}). The X-ray luminosity ($L_{\rm X}$)
was then calculated based on a distance equal to that of the
target. The distributions of minimum detectable $L_{\rm X}$ for both
samples are shown in Figure \ref{fig3}.

The latest spectral type companion to which the RASS observations are
sensitive to depends on the age of the target, derived from
theoretical isochrones (Fig. \ref{fig2} - \citealp{Marigo:2008p2100}), and the X-ray
luminosity sensitivity of the observations (Fig. \ref{fig3}). The
distribution of this spectral type sensitivity is given in
Figure \ref{fig4}. Most of the targets within both samples
had RASS observations sensitive to M-type companions and above: 75\%
of the X-ray sample and 85\% of the control sample. Nearly all the
RASS observations were sensitive to K-type companions --
87\% of the X-ray and 93\% of the control sample -- and the few
remaining targets were sensitive to F- or G-type companions.

The targets within the X-ray detected sample were chosen based on the
presence of a {\sl ROSAT} Bright Source Catalogue
(\citealp{Voges:1999p197} -- BSC) or {\sl ROSAT} Faint Source
Catalogue (\citealp{Voges:2000p3325} -- FSC) source within 35$\arcsec$
of the {\sl Hipparcos} coordinate of each target. As noted in Table
\ref{tab:xbinary}, 51\% are from the Faint Source Catalogue. Previous
correlations between BSC sources and optical star catalogues
(e.g. \citealp{Huensch:1998p2050}) have typically used a maximum
offset of 90$\arcsec$ between the catalogue positions, to define an
X-ray source. The distribution of the offsets
between X-ray source position and {\sl Hipparcos} position is given in
Figure \ref{fig5} and we have applied a more stringent maximum offset
cut-off than previous studies, 35$\arcsec$. All
of the A-type stars within the X-ray sample were also identified as
X-ray stars in previous studies of X-ray detected A-type stars
(e.g. \citealp{Schroder:2007p864}).

\section{Observations}
\begin{table*}
\caption{Instruments}
\begin{tabular}{lcccccccc}
Telescope &Proposal ID&Dates&$N$&Filter&$\lambda/D$&Field of
view&Pixel Scale&North\\
\hline
AEOS& - &02-02-2002 -- 03-03-2003&101&$I_{\rm C}$&$0\farcs05$&$24\farcs6\times24\farcs6$&$0\farcs048\pm0\farcs003$&$0.0^{\circ}\pm1.0^{\circ}$\\
CFHT&2008A-C22&12-06-2008 -- 14-06-2008&14& $H$ (FeII)&$0\farcs09$&$35\farcs6\times35\farcs6$&$0\farcs035\pm0\farcs0001$&$-2.4^{\circ}\pm0.1^{\circ}$\\
&2008A-C22&13-06-2008 -- 14-06-2008&1&$K^{\prime}$ (H$2_{1-0}$)&$0\farcs13$&$35\farcs6\times35\farcs6$&$0\farcs035\pm0\farcs0001$&$-2.4^{\circ}\pm0.1^{\circ}$\\
&2009B-C06&30-08-2009 -- 01-09-2009&18& $K^{\prime}$ (H$2_{1-0}$)&$0\farcs13$&$35\farcs6\times35\farcs6$&$0\farcs035\pm0\farcs0001$&$-2.4^{\circ}\pm0.1^{\circ}$\\
&2010A-C14&04-02-2010 -- 05-02-2010&29& $K^{\prime}$ (H$2_{1-0}$)&$0\farcs13$&$35\farcs6\times35\farcs6$&$0\farcs035\pm0\farcs0001$&$-2.4^{\circ}\pm0.1^{\circ}$\\
Gemini&GN-2008A-Q-74&18-06-2008 -- 24-06-2008&8& $K^{\prime}$ (Br$\gamma$)&$0\farcs06$&$21\farcs7\times21\farcs7$&$0\farcs021\pm0\farcs0001$&$0.5^{\circ}\pm0.3^{\circ}$\\
&GN-2008B-Q-119&17-08-2008 -- 25-11-2008&26& $K^{\prime}$ (Br$\gamma$)&$0\farcs06$&$21\farcs7\times21\farcs7$&$0\farcs021\pm0\farcs0001$&$0.5^{\circ}\pm0.3^{\circ}$\\
&GN-2009B-Q-120&08-09-2009 -- 19-12-2009&6& $K^{\prime}$ (Br$\gamma$)&$0\farcs06$&$21\farcs7\times21\farcs7$&$0\farcs021\pm0\farcs0001$&$0.5^{\circ}\pm0.3^{\circ}$\\
&GN-2010A-Q-75&10-05-2010&1& $K^{\prime}$ (Br$\gamma$)&$0\farcs06$&$27\farcs7\times21\farcs7$&$0\farcs021\pm0\farcs0001$&$0.5^{\circ}\pm0.3^{\circ}$\\
Palomar&-&11-04-2008 -- 13-07-2008&13& $H$ (CH$4_{\rm S}$)&$0\farcs07$&$25\farcs4\times25\farcs4$&$0\farcs025\pm0\farcs002$&$-0.7^{\circ}\pm0.1^{\circ}$\\
Palomar&-&11-04-2008 -- 13-07-2008&22& $K_{\rm S}$ (Br$\gamma$)&$0\farcs09$&$25\farcs4\times25\farcs4$&$0\farcs025\pm0\farcs002$&$-0.7^{\circ}\pm0.1^{\circ}$\\
\end{tabular}
\label{tab:observations}
\end{table*}
High resolution AO images were obtained for all 148 stars in order to
compare the binary statistics of the X-ray and control samples. The
data were acquired with several instruments listed in Table
\ref{tab:observations} -- VisIm \citep{Roberts:2002p2662} on AEOS, KIR
\citep{Doyon:1998p2790} on CFHT, NIRI \citep{Hodapp:2003p1692} on
Gemini North, and PHARO \citep{Hayward:2001p3209} on Palomar. The
resolution limit $\lambda$/$D$ ranged from 0$\farcs$05 for the $I_{\rm
  C}$-band AEOS images to 0$\farcs$13 for the $K^{\prime}$ CFHT
images. The filter used for observations with each instrument is
given in Table \ref{tab:observations}, alongside the corresponding
narrowband filter in parentheses. The FWHM of the image
cores typically matched the diffraction-limit, due to the high quality
AO correction on these bright stars. Given the
nearby distances of the targets ($D<170$ pc), the resolution limit
corresponds to projected separations of $\sim$10 - 20 AU. The
field-of-view ranges from 21$\farcs$7$\times$21$\farcs$7 to 
35$\farcs$6$\times$35$\farcs$6, making binary systems as wide as
$\sim$ 3000 AU detectable. The effective field-of-view for the
combined science images was increased by dithering the target on the
detector. The search range covers the peak of the binary separation distribution of lower mass stars
(e.g. \citealp{Duquennoy:1991p1342,Fischer:1992p2667}), important for
resolving the bulk of the binary population.

The observing strategy was consistent for all targets. To search for
close companions, unsaturated exposures were obtained of each target
using either a narrow-band or neutral-density filter. Exposure times
ranged from 0.01s to 4.0s, with stacks of 3 to 500 frames. To detect wider,
fainter objects approaching the bottom of the Main Sequence, longer exposures in a wide-band filter were recorded with total integration times
ranging from 41s to 720s. Details of the filter combinations are given
in Table \ref{tab:observations} and exposure times of individual
targets are listed in Tables \ref{tab:sample} and
\ref{tab:control}. Because of the brightness of the targets, all-sky
survey images from {\sl 2MASS} are saturated over a significant fraction of
the separation range covered by the images within this study.
\section{Data analysis}
The science images were processed with standard image reduction steps
including dark subtraction, flat fielding, interpolation over bad
pixels, and sky subtraction. Alignment of short exposure images was achieved through
Gaussian centroiding, while the saturated exposures were aligned by
cross-correlating the diffraction spikes
(e.g. \citealp{Lafreniere:2007p1668}). To improve the
measurable contrast ratios, a radial subtraction was performed on the
saturated images to suppress the seeing halo of the central
star. Finally, all the processed images were median-combined to
increase the signal-to-noise ratio of any detection

Candidates were identified by visual inspection, and the separation
and magnitude difference were measured for each candidate, as reported
in Tables \ref{tab:binary} and \ref{tab:xbinary}. The projected
separation between the central star and candidate was calculated from
the positions of the centroids of each component in the final
median-combined image. The uncertainty of the separation incorporates both
the uncertainty in the instrument pixel scale, given in Table
\ref{tab:observations}, and the standard deviation of the measurements
from each individual exposure. An estimate of the physical separation
in AU was then determined from the {\sl Hipparcos}-derived distance to the
primary. The position angle of each candidate was measured based on the
instrument field orientation, given in Table \ref{tab:observations}, and
the rotation angle on the sky for all Gemini and a subset of the AEOS
data. For data obtained at Palomar and CFHT, there is no instrument or
sky rotation. Typically, the total uncertainty is dominated by the measurement
uncertainty, however the lack of calibration measurements within some
of the observation runs requires a more conservative estimate of the
plate scale and angle of true north uncertainty.

The magnitude difference between each candidate and target star was
measured with aperture photometry. Using an aperture of twice the
FWHM, the fluxes for the candidate and unsaturated star were
measured. If the candidate was only detected in the saturated image,
then the comparison flux of the central star was scaled according to the
exposure time of the saturated image and the appropriate filter
bandpass. The reported magnitude difference uncertainty was estimated
as the standard deviation of the values from each processed image
before combination. Using the magnitudes of the target from the
{\sl Hipparcos} and {\sl 2MASS} \citep{Cutri:2003p2021} source catalogues, the apparent magnitude of the
candidate was determined. An estimate of the physical properties of both primary and candidate
companion was made using a combined set of theoretical
solar-metallicity isochrones
\citep{Marigo:2008p2100,Baraffe:1998p2099}. Each target was plotted on 
a colour-magnitude diagram (Fig. \ref{fig2}) from which an estimate of
the age was derived. Estimated colours and bolometric luminosities
were obtained for the companion candidates based on the measured
magnitude difference, using an isochrone of the same age as the primary. 

\section{Results}
\begin{table}
\caption{Candidate binary systems within control sample}
{\tiny \begin{tabular}{lccccc}
Designation&Separation&Position Angle&Magnitude&Filter&Observation\\
&\textit{arc sec}&\textit{degrees}&Difference&&Date\\
\hline
HIP2852 B$^{\dagger}$&0.93 $\pm$ 0.01&260.6 $\pm$ 0.3&5.07 $\pm$ 0.03&Br$\gamma$&17/10/2008\\
HIP9487 B&1.83 $\pm$ 0.01&266.9 $\pm$ 0.2&0.33 $\pm$ 0.01&H2$_{1-0}$&01/09/2009\\
HIP17572 B&3.4 $\pm$ 0.1&333.0 $\pm$ 1.0&2.54 $\pm$ 0.01&$I_{\rm C}$&04/02/2002\\
HIP28360 C&13.9 $\pm$ 0.3&155.0 $\pm$ 0.1&8.5 $\pm$ 0.2&$K^{\prime}$&05/02/2010\\
HIP29711 B&$\sim$4.2&$\sim$239.7&$<$ 2.5&$K^{\prime}$&25/11/2008\\
HIP35350 B&9.7 $\pm$ 0.1&33.8 $\pm$ 0.1&3.8 $\pm$ 0.1&Br$\gamma$&12/04/2008\\
HIP43570 B&0.66 $\pm$ 0.02&310.0 $\pm$ 1.0&2.58 $\pm$ 0.01&$I_{\rm C}$&04/02/2002\\
HIP44066 B&10.3 $\pm$ 0.3&320.9 $\pm$ 1.0&5.5 $\pm$ 0.2&$I_{\rm C}$&01/03/2003\\
HIP44901 B$^{\dagger}$&26.2 $\pm$ 0.1&33.9 $\pm$ 0.1&6.0 $\pm$ 0.1&$K^{\prime}$&05/02/2010\\
HIP51658 B&16.9 $\pm$ 0.04&357.6 $\pm$ 0.1&6.0 $\pm$ 0.2&$K^{\prime}$&04/02/2010\\
HIP54136 B&7.7 $\pm$ 0.3&110.7 $\pm$ 1.0&4.6 $\pm$ 0.2&$I_{\rm C}$&03/03/2003\\
HIP58510 B$^{\dagger}$&3.2 $\pm$ 0.1&218.4 $\pm$ 1.0&9.2 $\pm$ 0.3&$I_{\rm C}$&02/03/2003\\
HIP68520 Aa$^{\dagger}$&14.4 $\pm$ 0.5&41.9 $\pm$ 1.0&7.7 $\pm$ 0.1&$I_{\rm C}$&03/03/2003\\
HIP69592 B$^{\dagger}$&4.05 $\pm$ 0.03&174.5 $\pm$ 0.1&5.1 $\pm$ 0.1&CH4$_{\rm S}$&12/07/2008\\
HIP75043 B&0.26 $\pm$ 0.01&227.6 $\pm$ 2.0&6.0 $\pm$ 0.4&$I_{\rm C}$&29/05/2002\\
HIP84012 B&0.58 $\pm$ 0.01&236.0 $\pm$ 0.2&0.6 $\pm$ 0.1&Br$\gamma$&12/04/2008\\
HIP95081 B$^{\dagger}$&13.1 $\pm$ 0.1&16.9 $\pm$ 0.3&8.7 $\pm$ 0.1&$K^{\prime}$&24/06/2008\\
HIP101300 B&0.26 $\pm$ 0.01&241.7 $\pm$ 1.3&1.0 $\pm$ 0.1&$I_{\rm C}$&31/05/2002\\
HIP109667 B$^{\dagger}$&1.12 $\pm$ 0.01&285.2 $\pm$ 0.3&4.1 $\pm$ 0.1&Br$\gamma$&10/09/2008\\
&1.11 $\pm$ 0.01&284.7 $\pm$ 0.2&4.2 $\pm$ 0.1&H2$_{1-0}$&31/08/2009\\
\multicolumn{4}{l}{$\dagger$ - Previously unresolved companion candidate}
\label{tab:binary}
\end{tabular}}
\end{table}
\begin{table*}
\caption{Candidate binary systems within X-ray detected sample}
{\tiny \begin{tabular}{lccccccc}
Designation&Separation&Position Angle&Magnitude&Filters&Observation&Estimated&Estimated\\
&\textit{arc sec}&\textit{degrees}&Difference&&Date&$V-I$&$\log(L_{\rm
  X}/L_{\rm Bol})$\\
\hline
5310 B$^{\dagger\star}$&0.36 $\pm$ 0.01&175.3 $\pm$ 0.3&3.91 $\pm$ 0.04&Br$\gamma$&16/10/2008&1.98&-3.01\\
9480 B$^{\star}$&0.67 $\pm$ 0.01&297.3 $\pm$ 0.2&1.18 $\pm$ 0.02&H2$_{1-0}$&01/09/2008&0.64&-4.24\\
11569 B$^{\star}$&2.77 $\pm$ 0.01&230.0 $\pm$ 0.2&1.60 $\pm$ 0.02&H2$_{1-0}$&05/02/2010&0.64&-4.32\\
11569 C&7.22 $\pm$ 0.01&115.3 $\pm$ 0.1&1.98 $\pm$ 0.01&H2$_{1-0}$&05/02/2010&0.75&-4.08\\
13133 C$^{\dagger}$&$\sim$6.6&$\sim$70.5&$<1.9$&$K^{\prime}$&14/11/2008&0.86&-3.00\\
13133 D$^{\dagger\star}$&3.87 $\pm$ 0.03&229.8 $\pm$ 0.3&9.4 $\pm$ 0.3&$K^{\prime}$&14/11/2008&4.94&1.16\\
17608 Ab$^{\dagger}$&0.25 $\pm$ 0.02&111.0 $\pm$ 1.1&4.0 $\pm$ 0.4&$I_{\rm C}$&04/02/2002&0.54&-4.32\\
17923 B&3.1 $\pm$ 0.1&232.3 $\pm$ 1.0&2.66 $\pm$ 0.01&$I_{\rm C}$&03/02/2002&0.65&-3.29\\
17923 Ca&9.7 $\pm$ 0.3&233.7 $\pm$ 1.0&2.8 $\pm$ 0.1&$I_{\rm C}$&03/02/2002&0.68&-3.23\\
17923 Cb&10.2 $\pm$ 0.3&235.0 $\pm$ 1.0&4.5 $\pm$ 0.2&$I_{\rm C}$&03/02/2002&1.08&-2.52\\
19949 B$^{\dagger}$&13.5 $\pm$ 0.4&146.9 $\pm$ 1.0&7.3 $\pm$ 0.3&$I_{\rm C}$&05/02/2002&1.88&-2.63\\
20648 B&1.7 $\pm$ 0.1&337.9 $\pm$ 1.0&3.12 $\pm$ 0.02&$I_{\rm C}$&04/02/2002&0.80&-4.84\\
&1.80 $\pm$ 0.01&341.4 $\pm$ 0.1&2.55 $\pm$ 0.01&H2$_{1-0}$&04/02/2010&&\\
22287 Ab$^{\star}$&0.46 $\pm$ 0.01&41.6 $\pm$ 1.2&3.8 $\pm$ 0.1&$I_{\rm C}$&05/02/2002&1.13&-4.41\\
22287 B$^{\star}$&13.2 $\pm$ 0.5&238.7 $\pm$ 1.0&5.5 $\pm$ 0.2&$I_{\rm C}$&05/02/2002&1.82&-3.78\\
23179 B$^{\star}$&$\sim$4.7&$\sim$4.0&$<2.6$&$K^{\prime}$&15/11/2008&1.12&-2.81\\
24019 B$^{\star}$&$\sim$11.2&$\sim$26.1&$<2.4$&$I_{\rm C}$&04/02/2002&0.86&-4.38\\
28614 BaBb$^{\star}$&0.40 $\pm$ 0.01&22.1 $\pm$ 0.3&1.27 $\pm$ 0.01&$K^{\prime}$&19/12/2009&0.27&-5.36\\
29997 B$^{\dagger}$&8.47 $\pm$ 0.05&218.1 $\pm$ 0.4&6.8 $\pm$ 0.1&$K^{\prime}$&01/09/2009&2.43&-2.16\\
30419 B$^{\star}$&12.20 $\pm$ 0.04&29.0 $\pm$ 0.2&1.72 $\pm$ 0.03&H2$_{1-0}$&01/09/2009&0.65&-4.95\\
39095 B$^{\star}$&5.3 $\pm$ 0.2&65.8 $\pm$ 1.0&9.5 $\pm$ 0.7&$I_{\rm C}$&02/02/2002&2.32&-2.63\\
39847 Aa$^{\dagger}$&4.6 $\pm$ 0.2&114.0 $\pm$ 1.0&9.5 $\pm$ 0.5&$I_{\rm C}$&02/02/2002&2.39&-2.26\\
42313 Ab$^{\dagger}$&2.6 $\pm$ 0.1&265.1 $\pm$ 1.0&6.6 $\pm$ 0.2&$I_{\rm C}$&04/02/2002&1.79&-3.08\\
&2.6 $\pm$ 0.1&262.7 $\pm$ 3.1&7.3 $\pm$ 0.2&$I_{\rm C}$&01/03/2003&&\\
44127 B$^{\star}$&2.35 $\pm$ 0.02&76.6 $\pm$ 0.1&4.22 $\pm$ 0.02&Br$\gamma$&12/04/2008&2.08&-3.55\\
&2.40 $\pm$ 0.01&78.8 $\pm$ 0.1&4.36 $\pm$ 0.02&H2$_{1-0}$&05/02/2010&&\\
44127 C$^{\star}$&1.94 $\pm$ 0.02&79.8 $\pm$ 0.1&4.26 $\pm$ 0.02&Br$\gamma$&12/04/2008&2.09&-3.53\\
&1.92 $\pm$ 0.01&87.2 $\pm$ 0.1&4.30 $\pm$ 0.02&H2$_{1-0}$&05/02/2010&&\\
45688 B$^{\star}$&2.5 $\pm$ 0.1&222.6 $\pm$ 1.0&1.64 $\pm$ 0.03&$I_{\rm C}$&06/02/2002&0.48&-4.79\\
&2.60 $\pm$ 0.02&224.0 $\pm$ 0.1&1.2 $\pm$ 0.1&Br$\gamma$&12/04/2008&&\\
51200 B$^{\star}$&2.41 $\pm$ 0.01&304.1 $\pm$ 0.1&3.08 $\pm$ 0.01&H2$_{1-0}$&04/02/2010&1.42&-3.63\\
52913 B$^{\star}$&2.3 $\pm$ 0.1&13.8 $\pm$ 1.1&0.31 $\pm$ 0.01&$I_{\rm C}$&02/03/2003&0.31&-5.13\\
62394 Ab$^{\dagger\star}$&3.2 $\pm$ 0.1&348.8 $\pm$ 1.4&7.0 $\pm$ 0.2&$I_{\rm C}$&29/05/2002&2.16&-2.37\\
65241 B$^{\dagger\star}$&0.34 $\pm$ 0.02&41.7 $\pm$ 3.2&4.3 $\pm$ 0.4&$I_{\rm C}$&03/03/2003&1.40&-3.16\\
65477 B$^{\star}$&1.07 $\pm$ 0.01&209.0 $\pm$ 0.1&5.6 $\pm$ 0.1&CH4$_{\rm S}$&11/04/2008&2.19&-3.46\\
66249 B&1.81 $\pm$ 0.01&154.4 $\pm$ 0.1&6.4 $\pm$ 0.1&H2$_{1-0}$&05/02/2010&2.45&-3.18\\
66727 B&4.4 $\pm$ 0.1&338.5 $\pm$ 1.2&2.95 $\pm$ 0.02&$I_{\rm C}$&03/03/2003&0.74&-3.72\\
76376 C$^{\dagger\star}$&9.6 $\pm$ 0.4&350.1 $\pm$ 1.0&11.9 $\pm$ 0.3&$I_{\rm C}$&29/05/2002&3.6&-1.52\\
76878 B$^{\star}$&2.3 $\pm$ 0.1&53.4 $\pm$ 1.7&7.3 $\pm$ 0.2&$I_{\rm C}$&29/05/2002&2.24&-2.31\\
&2.4 $\pm$ 0.02&86.4 $\pm$ 0.1&5.1 $\pm$ 0.4&$K_{\rm S}$&13/07/2008&&\\
80628 B$^{\star}$&0.67 $\pm$ 0.01&22.6 $\pm$ 0.1&2.26 $\pm$ 0.03&Br$\gamma$&12/04/2008&0.90&-3.78\\
82321 B&1.82 $\pm$ 0.01&34.1 $\pm$ 0.1&2.2 $\pm$ 0.1&CH4$_{\rm S}$&12/07/2008&0.77&-4.66\\
82321 C&2.06 $\pm$ 0.02&38.4 $\pm$ 0.1&2.69 $\pm$ 0.03&CH4$_{\rm S}$&12/07/2008&0.90&-4.39\\
87045 B$^{\star}$&0.32 $\pm$  0.01&144.1 $\pm$ 1.1&2.45 $\pm$ 0.04&$I_{\rm C}$&29/05/2002&0.62&-3.71\\
88771 B$^{\star}$&24.83 $\pm$ 0.06&297.6 $\pm$ 0.1&5.2 $\pm$ 0.1&$K^{\prime}$&05/02/2010&2.18&-3.83\\
88771 D&24.20 $\pm$ 0.06&48.4 $\pm$ 0.1&8.1 $\pm$ 0.1&$K^{\prime}$&05/02/2010&3.40&-2.53\\
91971 B&23.28 $\pm$ 0.07&51.3 $\pm$ 0.2&9.7 $\pm$ 0.2&$K^{\prime}$&13/06/2008&4.27&-0.33\\
93747 B&7.27 $\pm$ 0.02&47.0 $\pm$ 0.2&4.88 $\pm$ 0.02&FeII&13/06/2008&1.99&-3.23\\
98103 C$^{\dagger\star}$&2.8 $\pm$ 0.1&184.7 $\pm$ 0.2&4.7 $\pm$ 0.1&$K^{\prime}$&18/06/2008&2.05&-2.96\\
102033 B$^{\star}$&0.72 $\pm$ 0.02&345.4 $\pm$ 1.1&2.36 $\pm$ 0.01&$I_{\rm C}$&31/05/2002&0.72&-4.62\\
106711 B$^{\dagger}$&6.98 $\pm$ 0.04&58.1 $\pm$ 0.3&8.5 $\pm$ 0.1&$K^{\prime}$&08/09/2008&3.00&-2.14\\
109521 B$^{\dagger\star}$&9.98 $\pm$ 0.06&241.3 $\pm$ 0.3&8.1 $\pm$ 0.1&$K^{\prime}$&08/09/2008&3.27&-1.18\\
110787 B$^{\dagger\star}$&0.29 $\pm$ 0.01&211.1 $\pm$ 0.6&4.3 $\pm$ 0.1&Br$\gamma$&17/09/2008&2.03&-3.18\\
117452 Ba$^{\dagger\star}$&3.7 $\pm$ 0.1&237.3 $\pm$ 0.4&3.48 $\pm$ 0.04&H2$_{1-0}$&30/08/2009&1.61&-3.01\\
117452 Bb$^{\dagger\star}$&3.5 $\pm$ 0.1&238.5 $\pm$ 0.5&3.7 $\pm$ 0.1&H2$_{1-0}$&30/08/2009&1.73&-2.90\\
\multicolumn{4}{l}{$\dagger$ - Previously unresolved companion candidate}\\
\multicolumn{4}{l}{$\star$ - Companion candidate falls within RASS error ellipse}
\label{tab:xbinary}
\end{tabular}}
\end{table*}

\begin{table*}
\caption{Results Summary}
\begin{tabular}{lcccc}
&\multicolumn{2}{c}{\textit{Total
    Field-of-View}}&\multicolumn{2}{c}{\textit{RASS Error Ellipse Search Area}}\\
&X-Ray&Control&X-Ray&Control\\
\hline
A8 -- M9 Companion\footnotemark&$60^{+6}_{-6}\%$&$20^{+5}_{-4}\%$&$43^{+6}_{-6}\%$&$12^{+4}_{-3}\%$\\
B6 -- A7 Companion&$2^{+4}_{-1}\%$&$2^{+3}_{-1}\%$&$3^{+4}_{-1}\%$&$2^{+3}_{-1}\%$\\
No Resolved
Companion&$38^{+6}_{-6}\%$&$78^{+4}_{-5}\%$&$56^{+6}_{-6}\%$&$86^{+3}_{-4}\%$\\
\label{tab:summary}
\end{tabular}
\\
\footnotesize $^1$ - Expected spectral type based on measured magnitude difference and
assuming the same distance as the target.
\end{table*}
\begin{figure}
\resizebox{\hsize}{!}{{\includegraphics{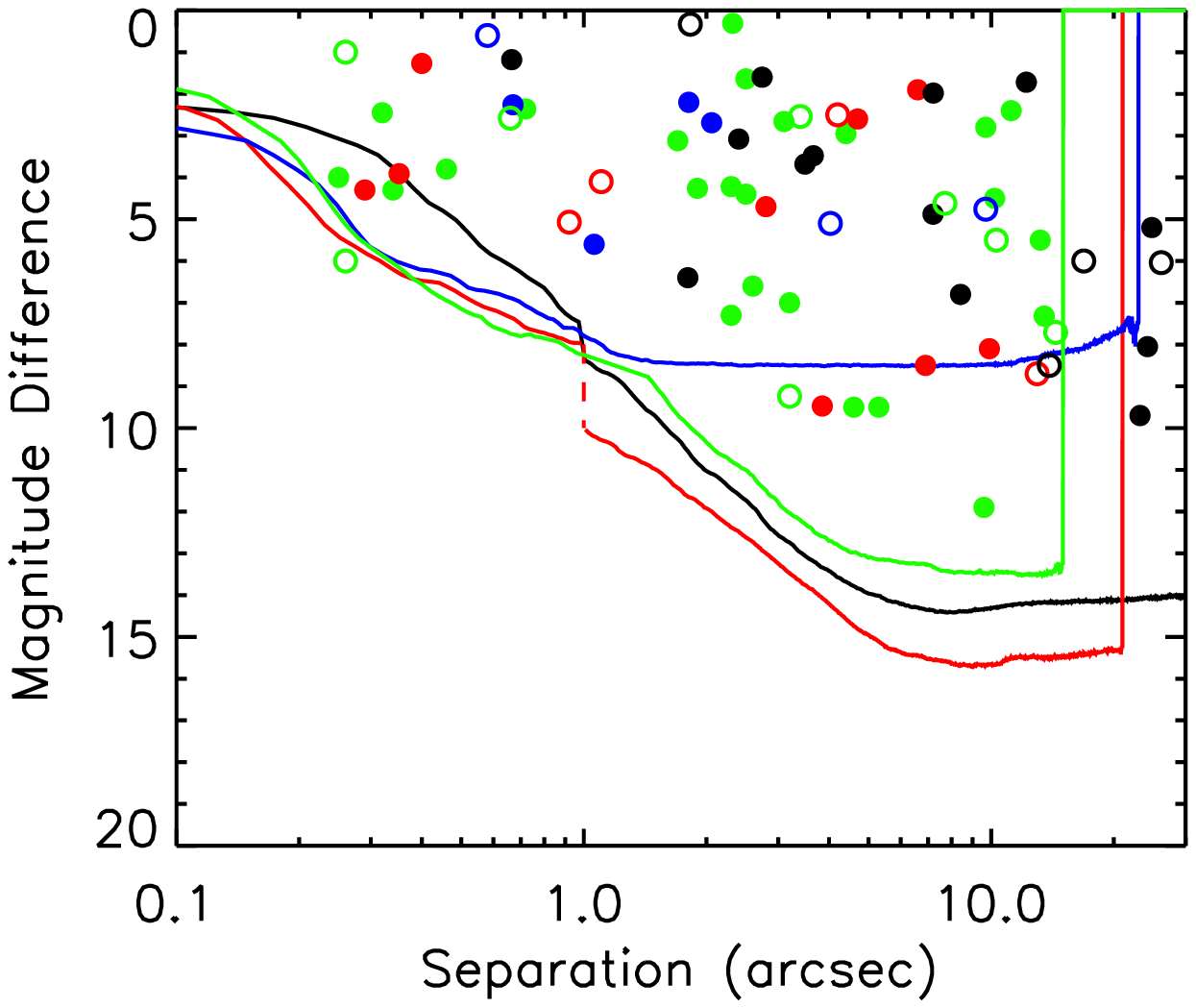}}}
\caption{The magnitude difference of the candidate companions detected
within this study as a function of angular separation from the central
star. Filled and open circles represent companions within the X-ray
and control samples, respectively. Colours represent each of the
instruments used: AEOS (green), CFHT (black), Gemini (red), Palomar
(blue). Over-plotted are the detection limits for each instrument (see
\S \ref{sec:limits}). The dashed portion of the Gemini sensitivity
curve represents the edge of the field of view for the unsaturated exposures.}
\label{fig6}
\end{figure}
\subsection{Detections}
Among the 148 targets, a total of 68 candidate companions were imaged
around 59 members of the total sample. One-third of the candidate
companions, 23 systems, are newly resolved.  The binary angular
separations range from 0$\farcs$3 to 26$\farcs$2, and the magnitude
differences range from 0.3 to 11.9, corresponding to spectral types of
mid-A to late-M for associated companions. The measured magnitude difference of
the candidates is plotted as a function of separation in Figure
\ref{fig6}. Properties of the companion candidates in the X-ray
and control samples are listed in Tables \ref{tab:binary} and
\ref{tab:xbinary}, respectively. Candidate companions are limited to
those with less than 5\% probability of being a background object,
based on the star density analysis described in \S\ref{sec:prob}. 
\subsection{Detection limits}
\label{sec:limits}
The sensitivity to companions varies with angular separation from
the central star due to the significant residual halo from the bright
targets. Detection limits for each image are quantified by determining
the flux level in a 5$\times$5 pixel aperture that would produce a
signal 5$\sigma$ above the noise within the aperture. The median
magnitude difference sensitivity curve for each instrument is plotted
in Figure \ref{fig6}. Since the data were obtained at several
wavelengths, the bottom of the Main Sequence corresponds to a
different magnitude difference for each instrument. For an A0 primary,
a companion at the bottom of the Main Sequence would have an absolute magnitude
of 14.3 at $I_{\rm C}$, 10.5 at $H$, and 10.2 at $K_{\rm S}$ at an age of 700 Myr. The infrared
data obtained at CFHT and Gemini are sensitive to the bottom of the
Main Sequence at separations beyond $\sim2\arcsec$. The achieved
contrast for the Palomar data was less due to the shorter exposure
times, and reached a companion mass limit of 0.12 M$_{\odot}$ to 0.2
M$_{\odot}$, depending on the age of the target. The AEOS data have a
sensitivity limit to companions ranging from 0.08 M$_{\odot}$ to 0.1
M$_{\odot}$. The sensitivity to companions for both the X-ray and
control samples is similar, making the difference between the two
measured binary frequencies a valid test of the companion hypothesis.
\subsection{Probability of chance superpositions}
\label{sec:prob}
\begin{figure}
\resizebox{\hsize}{!}{{\includegraphics{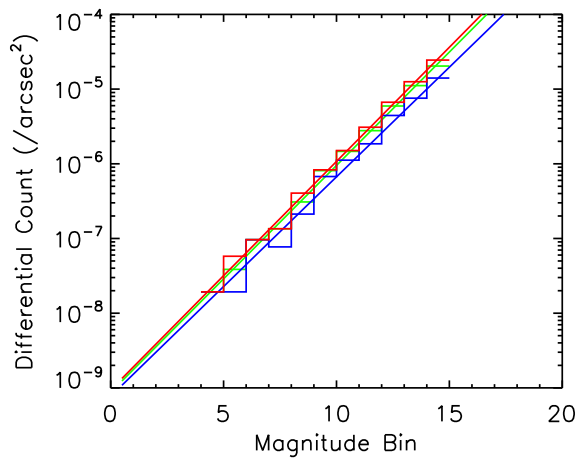}}}
\caption{Histogram of differential source counts and
  corresponding logarithmic fits within the {\sl 2MASS}
  source catalogue in the vicinity of HIP57646 in $J$, $H$ and
  $K_{\rm S}$ filters (blue, green and red respectively).}
\label{fig7}
\end{figure}
An estimate of the probability of each companion candidate
being an optical binary was made based on the local stellar
densities for each target, measured from the {\sl 2MASS} source catalogue. The number
of sources within a $2^{\circ}\times2^{\circ}$ box of each target was determined in
magnitude bins 1 magnitude in width from 0--14 mag
for the $J$, $H$, and $K_{\rm S}$ bandpasses. An example
plot of this differential source count per area is given in Figure
\ref{fig7}. A power law fit was applied to the counts such that
\begin{equation}
N=\pi\rho^210^{b+am}
\end{equation}
where $N$ is the number of sources within a separation $\rho$ from the
target, with an
apparent magnitude brighter than $m$, expressed as a function of the
two fit parameters $a$, the gradient, and $b$, the intercept. For the
$I_{\rm C}$ band observations obtained at the AEOS, we have approximated the local
stellar density using the $J$ band {\sl 2MASS} data. Candidates with
$N>0.05$ were assumed to be a background object, and not counted
for any aspect of this study -- a total of 492 candidates were rejected
through this process. To compare the stellar density across
the samples, Tables \ref{tab:sample} \& \ref{tab:control} give the
number of objects brighter than 14$^{\rm th}$ magnitude expected
per square arcminute in the vicinity of each target. In order to prove physical
association of the companion candidates which satisfy this criterion,
a second epoch measurement will be required.
\section{Discussion}
\subsection{Multiplicity comparison}
The frequency of multiple systems in the X-ray and control sample was
determined by two methods. In the first calculation, the total
field-of-view of each observation was used, and, in the
second calculation, the search area was restricted to the RASS
position error box. For each approach, candidate companions with a
small magnitude difference, consistent with a spectral type in the
B6-A7 range, were excluded from the X-ray companion hypothesis
  test and are listed separately in Table \ref{tab:summary}. This
  criterion of a companion capable of generating X-rays eliminated one
  binary companion from the X-ray sample and two companions from the control sample. All
  multiple systems considered also satisfied the background object
  probability of $<$ 5\%, as described in \S\ref{sec:prob}.

Considering the total field-of-view of the combined dithered
observations, candidates satisfying the magnitude and background
probability criteria were included in the multiple frequency
measurement. Among the X-ray sample, $60^{+6}_{-6}\%$ were
multiple, compared to $20^{+5}_{-4}\%$ for the control
sample -- a difference of $40\pm8\%$, a 5$\sigma$ result. These and
subsequent reported errors are estimated from a
binomial distribution (e.g. \citealp{Burgasser:2003p3394}). Spectroscopic
binaries -- unresolved with these observations -- constitute a
significant fraction of both samples ($\sim$ 15\%,
\citealp{Pourbaix:2004p2770}). This estimate represents a lower limit
on the frequency since the sample of stars observed with the radial
velocity monitoring is not known, and the large $v\sin i$ of the
primary and less massive unseen companions make such observations
challenging. These spectroscopically resolved binaries are not
considered within our statistics.

The multiplicity of the X-ray sample was also measured by considering
only companion candidates that were located within the confines of the
RASS error ellipse. For each target, the AO data covered a
  portion of the RASS error ellipse ranging from 25 to 100 percent. This additional restriction
lowered the multiple frequency to $43^{+6}_{-6}\%$. To determine a
comparable frequency for the control sample, a series of companion
searches were performed by randomly assigning the RASS-optical offset
and corresponding error ellipse of an X-ray target to a control target
and determining the number of candidate companions which fall within
the error ellipse. Based on a large number of simulations (100,000),
the frequency of multiples was estimated as
$12^{+4}_{-3}\%$. These two
frequencies are different by $31\pm7\%$, a 4$\sigma$ result.

A summary of the multiplicity calculations is given in Table \ref{tab:summary}. The
high statistical significance of the difference in frequencies for the
X-ray and control samples provides strong support of the companion
hypothesis as an explanation of the X-ray detection of B6-A7
stars. Further evidence for individual systems with separations of a
few arcsecond could be provided by high resolution {\sl Chandra}
observations which would have the pointing accuracy to assign the
X-ray flux to the companion unambiguously. One target, Merope in
the Pleiades, was observed with the high resolution mode of {\sl
  Chandra}, but the binary separation is only 0$\farcs$25, making the
discrepancy between the {\sl Chandra} and {\sl 2MASS} coordinates ambiguous
in this case. Targets within the X-ray sample for which no companions have been
resolved will make prime targets for future interferometric and
spectroscopic study in a search for lower-mass companions with angular
separations low enough to render them undetectable with AO
observations.

\subsection{{\sl ROSAT} positional uncertainty}

Previous studies of the unexplained X-ray detection of early-type
stars (e.g. \citealp{Schroder:2007p864}) have used the same definition
of an X-ray detected early-type star as presented by
\citet{Huensch:1998p2050} -- any X-ray source within $90\arcsec$ of an
optical source can be attributed to the optical source. This value was
based on estimating the frequency of false attribution by means of a
Monte Carlo simulation, and was selected at the radius at which the
probability of correctly attributing an X-ray source is $\sim$50 percent. A
significantly lower offset of $25\arcsec$ was calculated by
\citet{Voges:1999p197} from a correlation of the {\sl Tycho} catalogue
and {\sl ROSAT} Bright Source Catalogue positions, a radius within
which 90 percent of the optical targets have an X-ray source
attributed. This measurement represents the empirical positional
uncertainty of the RASS source catalogue positions.

The sample investigated within this study was initially selected in
the same manner as \citet{Huensch:1998p2050} -- using a maximum offset
of $90\arcsec$. The tail of the offset distribution was removed by
applying a more stringent maximum offset at $35\arcsec$, as described
in \S \ref{sec:sample}. Variations in the field-of-view size between
instruments caused the coordinates of the X-ray source given within
the RASS to be outside of the field-of-view within a small subset of
the observations. In order to investigate any biases this may have had
upon the results presented previously, the sample was further
restricted to only include those targets for which the RASS source
position was within the field-of-view and at least 50 percent of the
RASS error ellipse was covered -- a total of 45 stars. For this sample
a marginally higher frequency of companions located within the RASS
error ellipse was recorded, $53^{+7}_{-7}\%$, reinforcing the
result obtained with the unrestricted sample.

\subsection{Comparison of measured and expected X-ray luminosities}

\subsubsection{Candidate companions with measured colours}
\begin{figure}
\resizebox{\hsize}{!}{{\includegraphics{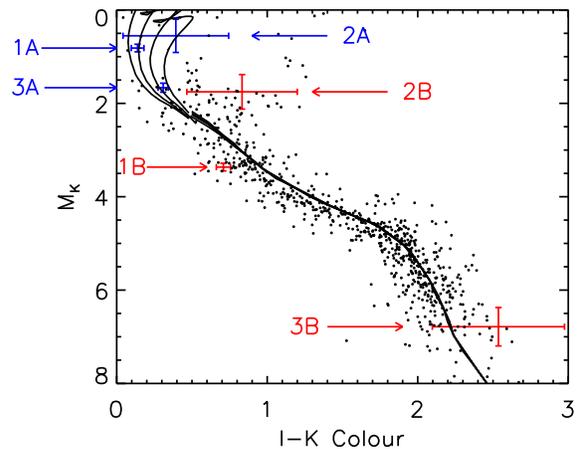}}}
\caption{A colour-magnitude diagram of 792 nearby Gliese
  stars. Overplotted are four theoretical isochrones of ages $\log
  t=8.7, 8.8, 8.9, 9.0$
  \citep{Marigo:2008p2100,Baraffe:1998p2099}. Three of the targets
  within the X-ray sample are plotted in blue: HIP 20648 (1), HIP 45688
  (2) and HIP 76878 (3). The corresponding resolved companions for each
  primary are shown in red.}
\label{fig8}
\end{figure}
\begin{figure*}
\resizebox{\hsize}{!}{{\includegraphics{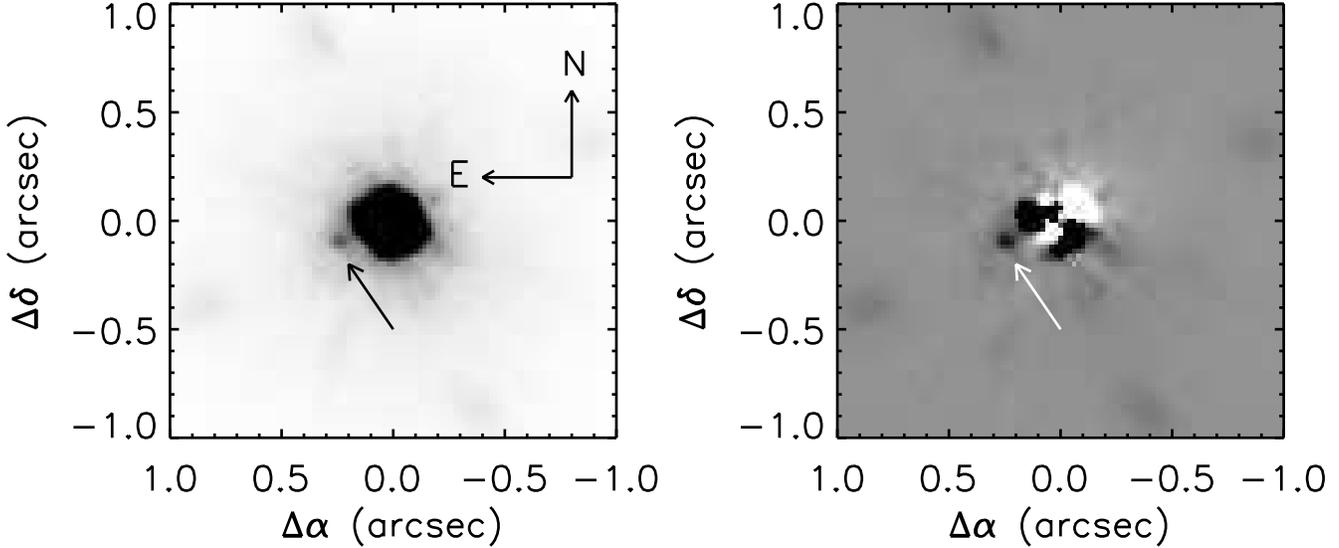}}}
\caption{A sub-arcsecond companion candidate is resolved
around HIP17608 (Merope), a member of the Pleiades cluster. A faint
($\Delta I=4.0 \pm 0.4$)
companion candidate at $\rho =0\farcs25$, $\theta =110^{\circ}$ is
visible within the median combined image of the 500, 0.048 second
unsaturated exposures (\textit{left panel}). The scale is linear from
0 (white) to 45 (black). After radial subtraction the object becomes
more prominent (\textit{right panel}), with a linear scale between -15
and 20.}
\label{fig9}
\end{figure*}
Several of the candidate companions to X-ray targets have a measured
$I-K$ colour from this study, and are plotted in Figure \ref{fig8}. The colour provides additional
information to estimate the spectral type of the object and to test
further the capacity of the second object to generate X-ray
emission.  The three systems
with colours are: (1) HIP 20648, (2) HIP 45688, and (3) HIP 76878. The
$I-K$ colours of the candidate companions are all consistent with 
X-ray emitting companions: 0.71 $\pm$ 0.05 or late F-/early G-type for HIP 20648 B, 0.83
$\pm$ 0.37  or mid G-type for HIP 45688 B, and 2.54 $\pm$ 0.44 or late M-type for
HIP 76878 B.

With the assumption of a distance and age equivalent to the primary
distance, the X-ray luminosity associated with the {\sl ROSAT} detection can
by checked for consistency with the spectral type. The position on the
colour-magnitude diagram for each
primary star and its imaged candidate companion is given in Figure
\ref{fig8}, assuming the distance to each 
component is the same. Each case is examined individually, and the 
colour and proper motion measurements clearly support the assignment of the
X-ray emission to the candidate companion in two cases, while one case
remains uncertain. 

The theoretical isochrone that best fits the
first target, HIP 20648A, corresponds an age of $\sim$650 Myr, and the
candidate companion position in Figure \ref{fig8} is as expected for an
associated companion. The companion X-ray luminosity is $\log L_{\rm
  X}=28.71$, and this value falls between the X-ray luminosities of 
Hyades F- and G-type stars. The assessment of the second target, HIP 45688,
is complicated by the presence of a known close companion to the
imaged candidate companion ($\rho \sim 0\farcs06$ -- \citealp{McAlister:1993p2768}), unresolved in the current data. The composite
colour and magnitude of the BaBb system appear to be more luminous than
expected for an object at the 630 Myr age estimated for the primary,
even if the pair is an equal 
magnitude binary. The X-ray luminosity of BaBb would be $\log L_{\rm
  X}=29.43$, significantly higher than younger G-type stars in the Hyades. 
For the final system, HIP 76878, the best fit age is 700 Myr, similar to the Hyades.
The X-ray luminosity of the candidate companion is
$\log L_{\rm   X}=29.26$, if the distance is equal to that of
HIP76878. This X-ray level is
higher than observed X-ray luminosities of M-dwarfs of similar age
within the Hyades \citep{Stern:1995p2131}. In this case, the time
baseline between the two observations also reveals a significant
motion of the candidate relative to the primary on a trajectory
different from both a background object and a bound companion. The presence of a
foreground M-dwarf in a chance superposition with HIP76878 explains
this discrepant proper motion, the red colour of the object, and the
unusually high X-ray luminosity.

\subsubsection{Candidate companions in open clusters}
A subset of the X-ray detected targets with imaged candidate
companions are members of
stellar clusters. HIP 17608 and HIP 17923 are Pleiades members,
while HIP 20648 is a Hyades member. Extensive X-ray population studies
of both the Pleiades
(e.g. \citealp{Micela:1985p1704,Stauffer:1994p2737,Daniel:2002p1719})
and the Hyades (e.g. \citealp{Micela:1988p2791,Stern:1995p2131}) have been
conducted with {\sl Einstein} {\sl ROSAT} and {\sl Chandra}, providing comparison 
X-ray luminosities to test the likelihood that the candidate
companions are responsible for the detected X-ray emission. 

The candidate companion to HIP 17608 (Merope in the Pleiades) with
0$\farcs$25 separation is shown in
Figure \ref{fig9}. With a magnitude difference of $\Delta
I_{\rm C}=4.0\pm0.4$, the second object is a mid F-type
star if associated. Assuming a distance to the Pleiades of 133 pc
\citep{Pan:2004p2793}, the X-ray luminosity for the HIP 17608 system
is $\log L_{\rm X}=29.91$. The typical X-ray luminosity of F-dwarfs
within the Pleiades is estimated to be $\log L_{\rm
  X}\sim29.43\pm0.29$ \citep{Stauffer:1994p2737}, indicating that the
companion to HIP17608, if associated, is on the upper limit of X-ray
activity for this class of star. 

For the second Pleiades member, the observations resolve
three of the companions (B,Ca,Cb) within the HIP17923 
quintuple system. Based on the measured magnitude differences, we estimate
the mass of the components as follows: B -- 1.2 $\pm$ 0.1 M$_{\odot}$
(mid F-type), Ca -- 1.2 $\pm$ 0.1 M$_{\odot}$ (mid F-type), Cb --
0.9 $\pm$ 0.1 M$_{\odot}$ (mid G-type). Deeper X-ray observations of
the Pleiades \citep{Micela:1999p2176} revealed an estimated X-ray
luminosity of $\log L_{\rm X}=30.08$ for this
system. If the X-ray counts were distributed evenly
between the three later-type companions resolved within our AO images,
the individual X-ray luminosities would be $\log L_{\rm X} \sim$29.6,
similar to G- and F-type Pleiades members
\citep{Stauffer:1994p2737}.

The final cluster X-ray target with a
resolved companion is the Hyades member HIP 20648. As described in
\S 7.2, the candidate companion also has a measured I-K colour
consistent with a late F-/early G-type star, and the X-ray luminosity
assigned to the target is consistent with a Hyades G-type star
\citep{Stern:1995p2131}.

\subsubsection{Remaining candidate companions}
\begin{figure}
\resizebox{\hsize}{!}{{\includegraphics{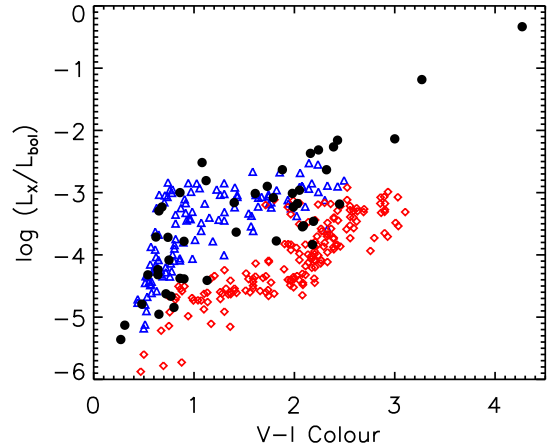}}}
\caption{ The ratio of X-ray to bolometric luminosity is plotted as a
  function of colour for the candidates resolved within this
  study. The majority of candidates are constrained by the Pleiades
  (blue triangles) and Hyades (red diamonds) members, representing the
two age extremes of the sample.}
\label{fig10}
\end{figure}

For the remaining candidate companions, an estimate of the ratio of
X-ray to bolometric luminosity can be made under the assumption that
the candidate is a physical companion at the same distance. From the
absolute magnitude, the $V-I$ colour  and bolometric luminosity were
inferred from theoretical isochrones \citep{Baraffe:1998p2099}. The
ratios of the observed X-ray luminosity to the estimated bolometric
luminosity ($L_{\rm X}/L_{\rm bol}$) are plotted as a function of
$V-I$ colour in Figure \ref{fig10}, with Pleiades and 
Hyades members \citep{Zuckerman:2004p3390} overplotted as reference
populations spanning the age range of the sample.

All but two of the candidates are within the region bound by the
$\sim100$ Myr Pleiades and $\sim650$ Myr Hyades members. Uncertainty
exists on both axes since both the $V-I$ colour and bolometric luminosity are
estimated from theoretical isochrones, assuming the distance. Future
observations to accurately determine the colour of these candidates will provide a more robust estimate of the bolometric
luminosity. The two outlying candidates shown in Figure \ref{fig10}
have unphysical high luminosity ratios, significantly higher than
the observed luminosity ratios of late M-type stars
(e.g. \citealp{Pizzolato:2003p2174}). In these two cases, additional
unresolved companions, or background X-ray sources, present a more feasible
explanation for the detected X-ray flux. The rate of false-detections,
$2/49$, corresponds to the 5\% contamination introduced through the
statistical method applied to the candidates to remove background
sources, as described in \S6.3.

\section{Summary}
In summary, a total of 148 stars with spectral types in the range
B6-A7 and distances of $<200$ pc have been observed with AO-equipped
cameras on 3.8m-8m telescopes. The high resolution images were
sensitive to companions with angular separations from $\sim$0$\farcs$3
to 26$\farcs$2 and magnitude differences extending to $\sim$14 mag. A
total of 68 candidate companions to 59 targets were resolved, and the
frequency of multiple systems was measured to be substantially higher
for the X-ray detected sample. The high frequency of multiples,
$43^{+6}_{-6}\%$, compared to $12^{+4}_{-3}\%$ for the
control sample is different by 4$\sigma$ and provides strong evidence
that the source of the X-ray emission is the candidate companion. The
X-ray detected stars with no resolved companion make ideal candidates
for future interferometric observations, as this study has shown that
the X-ray detection is indicative of the presence of an unresolved
companion, and interferometry can resolve binaries below the
resolution of the adaptive optics data presented here.

For three candidate companions to X-ray targets, the $I-K$ colour was
also measured, and the colours are consistent with late F- to late M-type stars,
supporting the identification of the second object as the X-ray
source in two cases. Among the X-ray targets with candidate companions, there are
also three cluster members, and the known age, distance, and cluster X-ray
properties enabled a further test of the companion X-ray luminosity
with other cluster members. In each case, the companions -- if
associated -- would have an X-ray luminosity similar to, or on the
upper range of, cluster stars with similar magnitude. Follow-up
observations of the X-ray targets with candidate companions using {\it
  Chandra} would provide the angular resolution in the X-ray band
necessary to confirm the second object as the true source of the X-ray
emission.

\section*{Acknowledgements}
We gratefully acknowledge several sources of funding. R. J. DR. and J. B (ST/F 007124/1) are funded through studentships
from  the Science and Technology Facilities Council (STFC). This work
was initiated with a grant awarded to J. P. from the Air Force Office
of Scientific Research (AFOSR) for the AEOS component and completed
with a grant from the STFC (ST/F003277/1). Funding for collaborative
visits was provided by the CONSTELLATION EC Research Training Network. Portions of this work were
performed under the auspices of the U.S. Department of Energy by
Lawrence Livermore National Laboratory in part under Contract
W-7405-Eng-48 and in part under Contract DE-AC52-07NA27344, and also
supported in part by the NSF Science and Technology CfAO, managed by
the UC Santa Cruz under cooperative agreement AST 98-76783. This work
was supported, through J. R. G., in part by University of California
Lab Research Program 09-LR-118057-GRAJ and NSF grant AST-0909188. We
thank LLNL summer students C. White (US Air Force Academy) and S. Kost
(Carnegie Mellon University) who assisted with obtaining a subset of
the data and some of the early analysis. This research has made use of the
SIMBAD and VizieR databases, operated at CDS, Strasbourg, France. This
publication makes use of data products from the Two Micron All Sky
Survey, which is a joint project of the University of Massachusetts
and the Infrared Processing and Analysis Center/California Institute
of Technology, funded by the National Aeronautics and Space
Administration and the National Science Foundation. This research has
made use of the Washington Double Star Catalog maintained at the
U.S. Naval Observatory. We thank the referee for the helpful comments
during the review process.
\bibliographystyle{mn2e}
\bibliography{Papers}
\label{lastpage}
\end{document}